\documentclass[aps,prd,reprint,groupedaddress,nofootinbib]{revtex4-2}
\usepackage{amsmath, amsfonts, amssymb, amsthm}
\usepackage{graphicx}

\newcommand\re {\mathrm{Re}}
\newcommand\im {\mathrm{Im}}

\newcommand\Renyi {R\'enyi\ }
\begin{document}


\title{A Numerical Calculation of Entanglement Entropy in de Sitter Space}

\author{Konstantinos Boutivas}
\email{kboutivas@phys.uoa.gr}
\affiliation{National and Kapodistrian University of Athens, Department of Physics, 15784 Zografou, Attiki, Greece}
\author{Dimitrios Katsinis}
\email{dkatsinis@phys.uoa.gr}
\affiliation{National and Kapodistrian University of Athens, Department of Physics, 15784 Zografou, Attiki, Greece}
\author{Georgios Pastras}
\altaffiliation[Also at ]{Laboratory for Manufacturing Systems and Automation, Department of Mechanical Engineering and Aeronautics, University of Patras, 26110 Patra, Greece}
\email{pastras@lms.mech.upatras.gr}
\affiliation{National and Kapodistrian University of Athens, Department of Physics, 15784 Zografou, Attiki, Greece}
\author{Nikolaos Tetradis}
\email{ntetrad@phys.uoa.gr}
\affiliation{National and Kapodistrian University of Athens, Department of Physics, 15784 Zografou, Attiki, Greece}


\date{\today}

\begin{abstract}
The entanglement entropy of a massless scalar field in de Sitter space depends on multiple scales, such as the radius of the entangling surface, the Hubble constant and the UV cutoff. We perform a high-precision numerical calculation using a lattice model in order to determine the dependence on these scales in the Bunch-Davies vacuum. We derive the leading de Sitter corrections to the flat-space entanglement entropy for subhorizon entangling radii. We analyze the structure of the finite-size effects and show that the contribution to the entanglement entropy of the sector of the theory with vanishing angular momentum depends logarithmically on the size of the overall system, which extends beyond the horizon.
\end{abstract}


\maketitle

\section{Introduction\label{sec:intro}}
The calculation by Srednicki of the entanglement entropy of a massless scalar field at its ground state in Minkowski space has led to the remarkable conclusion that it is dominated by an area-law term \cite{Srednicki:1993im} (see also \cite{Sorkin:1984kjy,Bombelli:1986rw}), bearing a strong resemblance to the entropy of black holes. This term is UV divergent. For a spherical entangling surface of area  $A_p$, it it proportional to $A_p/\epsilon^2_p$, with $\epsilon_p$ the UV cutoff. It has also been established that the first subleading term  is logarithmic, of the form $ \ln(A_p/\epsilon^2_p)$ \cite{Solodukhin:2008dh,Casini:2009sr,Lohmayer:2009sq}. The entanglement entropy of a massless field in de Sitter space was studied in \cite{Maldacena:2012xp}. It was parameterized as 
\begin{multline}\label{eq:SEE_expansion_mald}
S_{\textrm{dS}}=c_1\frac{A_p}{\epsilon_p^2}+\ln\left(H \epsilon_p\right)\left(c_2 + c_4 A_p H^2\right) \\ + c_5 A_p H^2 - \frac{c_6}{2} \ln\left(A_p H^2\right) +\textrm{const}.
\end{multline}
The Hubble constant $H$ is included in the arguments of the logarithms as a reference scale in order to make them dimensionless. In terms of the conformal time $\tau$, the scale factor is $a(\tau)=-1/(H \tau)$, with $\tau$ taking values between $-\infty$ and 0. The proper area $A_p$ is $A_p={A}/(H^2\tau^2)$, where $A$ is the area in comoving coordinates. The physical cutoff is related to the comoving one as $\epsilon_p=a \epsilon$. Similar relations apply to all other scales. 

The purpose of the current work is to examine in detail the dependence of the entanglement entropy on the various scales. We consider a massless scalar theory in the Bunch-Davies vacuum of de Sitter space. We employ a generalization of the method of Srednicki \cite{Srednicki:1993im}, which relies on the discretization of the theory. It has the advantage that it results in a well-defined, finite-dimensional eigenvalue problem that provides the spectrum of the reduced density matrix. On the other hand, this eigenvalue problem has to be solved numerically. Drawing conclusions for the continuous, infinite-size theory regarding terms beyond the dominant one is a formidable task. In the case of flat space such a calculation has been performed in \cite{Lohmayer:2009sq}. In \cite{Boutivas:2023ksg,Boutivas:2023mfg} we drew some basic qualitative conclusions, such as that entanglement entropy rises as time advances. We also found that a volume term appears in the eras of radiation and matter domination, in which the scale factor $a(\tau)$ grows $\sim \tau$ and $\sim \tau^2$, respectively, with $\tau>0$. Remarkably, such a term is absent in the inflationary era, in which $a(\tau)\sim -1/\tau$, with $\tau<0$.  However, we were unable to calculate coefficients such as those appearing in \eqref{eq:SEE_expansion_mald}. In this work we take on this task. We focus on a de Sitter background, in which the entanglement entropy has a more constrained structure.

In principle, the presence of  three dimensionful scales (the radius of the entangling surface $R_p$, the Hubble constant $H$ and the UV cutoff $\epsilon_p$) implies that the entaglement entropy could have a more general structure than that of equation \eqref{eq:SEE_expansion_mald}. In his seminal work \cite{Page:1993df} Page argued that the entanglement entropy of a quantum system at an arbitrary state is close to its maximal value, which is proportional to the number of degrees of freedom of the smaller subsystem.  The proportionality constant is the logarithm of the dimensionality of the Hilbert space of each local degree of freedom. In $3+1$ dimensions the number of degrees of freedom inside an entangling surface of characteristic scale $R_p$ is $\sim (R_p/\epsilon_p)^3$, where the UV cutoff $\epsilon_p$ is set by the lattice spacing. This implies that terms proportional to $\epsilon_p^k$, with $k<-3$ cannot emerge.

Another important property of entanglement entropy is the fact that, when the overall system lies in a pure state, the entropy of any subsystem is equal to that of its complement. For pure states, entanglement entropy is symmetric. Scalar field theory at the ground state is the most striking example, since in this case the entanglement entropy scales with the area of the subsystem \cite{Srednicki:1993im}. The entangling surface is literally the first common characteristic of the two complementary subsystems that comes to mind. However, for more general states, such as squeezed states, which are very similar to the states we consider in this work, the leading term is proportional to the volume of the smaller subsystem  \cite{Katsinis:2023hqn,Katsinis:2024sek}. 

The state we consider here corresponds to the Bunch-Davies vacuum of a quantum field theory in a de Sitter background. The wave function of each field mode can be interpreted as that of a squeezed state, but with a specific time-dependent squeezing parameter. As a result, the system shares features both with the vacuum and the squeezed states. In particular, the entanglement entropy has been observed to still obey an area law, while a volume term does not emerge \cite{Boutivas:2023ksg,Boutivas:2023mfg}. One of the aims of the current work is to reconfirm and solidify this conclusion through a precise numerical analysis. The main task is to identify as many new terms as possible in an expansion of the entanglement entropy such as \eqref{eq:SEE_expansion_mald}.

For scalar field theory at its ground state in flat space, entanglement entropy is insensitive to the IR. In other words one may define a finite-size system resulting from discretizing the theory on a finite lattice, calculate the entanglement entropy, and then remove the IR cutoff. For a massless theory in flat space, the dependence of the entropy on the IR regulator has been studied thoroughly in \cite{Lohmayer:2009sq}. In particular, it was shown that for each sector of fixed angular momentum number $\ell$ in an expansion in spherical harmonics, the leading finite-size correction scales as $N^{-2\left(\ell+1\right)}$, where $N$ is the number of the degrees of freedom of the sector. However, for squeezed states, such as those appearing in the case of de Sitter space, the various sectors display a stronger dependence on the size of the overall system. As we shall see in the following, contrary to the flat-space case, the entanglement entropy in de Sitter space actually depends on an IR cutoff that can be identified with the size of the overall system $L_p$. In our analysis we show that at early times, in the infinite-size limit, a term of the form $\ln\left(L_p/\epsilon_p\right)$ appears, originating in the $\ell = 0$ sector of the theory. 

Taking into account the above remarks, and thus allowing for a term proportional to the volume of the smaller subsystem, as well as a logarithmic term that depends on the size of the overall system, we may write the entanglement entropy in the form
\begin{multline}\label{eq:SEE_expansion}
S_{\textrm{dS}}= \frac{R_p^3}{\epsilon_p^3} f_3\left(H R_p\right) + \frac{R_p^2}{\epsilon_p^2} f_2\left(H R_p\right) + \frac{R_p}{\epsilon_p}  
f_1\left(H R_p\right)\\ +f_0\left(H R_p\right) + g\left(H R_p\right)\ln\frac{R_p}{\epsilon_p} + h\left(H R_p\right)\ln\frac{L_p}{\epsilon_p},
\end{multline} 
for a spherical subsystem of physical radius $R_p$, in the limit $\epsilon_p\ll R_p \ll L_p$. The functions $f_3$, $f_1$ and $h$ should vanish for $H\rightarrow 0$, while $f_2$, $f_0$ and $g$ should approach the constants of the flat-space case. Of those, the coefficient of the logarithmic term has a universal value equal to $-1/90$ \cite{Solodukhin:2008dh,Casini:2009sr,Lohmayer:2009sq}. In the context of the perturbative scheme that we shall employ in the following sections, the last term is always subleading to the flat-space entanglement entropy. The validity of this approximation sets constraints on the value of $H R_p$. In this sense, the above equation does not display an IR divergence for $L_p\to \infty$, since in this limit the expansion breaks down. Rather it demonstrates that there is an unexpected novel subleading term, induced by the expansion of the background, that probes the whole system, not just the interior of some subsystem. The presence of this term is confirmed both numerically and analytically as the leading correction to the flat-space result. 

Extracting the form of the various functions in \eqref{eq:SEE_expansion} through a numerical calculation, without some analytical guidance,  is very difficult. For this reason, we focus on  entangling surfaces with subhorizon radii, so that $H R_p = R/|\tau| \ll 1$, where $R=R_p/a$ is the comoving radius and we have made use of the relation $a(\tau)=1/(H|\tau|)$. In this limit, we can use a perturbative expansion in powers of $1/|\tau|$ around $\tau =-\infty$. Our assumption of a Bunch-Davies vacuum implies that the leading order is nothing but the flat-space limit. The subleading terms will involve only powers of $1/|\tau|$, while non-analytic terms, such as $\ln|\tau|$, would not be captured in this expansion. Such terms would be visible in an expansion in $|\tau|$ around $\tau=0$. Notice that the subleading nature of the last term in  \eqref{eq:SEE_expansion} is manifest in the $1/|\tau|$ expansion, since the leading term in the expansion of the function $h$ is $\sim R^2/\tau^2$. Naively taking $L_p\to \infty$ invalidates the expansion.  

We point out that \eqref{eq:SEE_expansion} is not the most general expression that one could write for the entropy. It is, however, adapted to an expansion in $HR_p$, which is tied to the early-time expansion. Notice that no explicit $\ln H \epsilon$ term appears in \eqref{eq:SEE_expansion}. It can be introduced by splitting the logarithm in the fifth term as $-\ln H \epsilon+ \ln HR_p$ and absorbing $g(HR_p)\ln HR_p$ in $f_0(HR_p)$. However, our analysis does not indicate the presence of a logarithmic term in $f_0(HR_p)$, as it is based on a regular expansion around the flat-space result at early times. Such a logarithmic term may be present in the late-time regime. In this sense our calculation is complementary to the one performed in \cite{Maldacena:2012xp}.

The structure of the paper is as follows: In section \ref{sec:review} we review the methodology for the calculation of entanglement entropy using a discretized version of the continuous field theory. In section \ref{sec:numerical_analysis} we focus on the technical details of the numerical calculation, concerning especially the identification and subtraction of finite-size effects, along with the summation of the contributions of the angular-momentum sectors. In section \ref{sec:results} we summarize the results of the analysis. In section \ref{conclusions} we present our conclusions. Finally, there is a series of appendices containing details about the C++ code we use and the analysis of the data.

\section{Methodology review\label{sec:review}}

Most calculations of entanglement entropy are performed in theories with very special symmetries, like conformal field theories. The replica-trick method is often applied \cite{Calabrese:2004eu,Calabrese:2009qy}, or the calculation is performed in the context of the AdS/CFT correspondence via the Ryu-Takayanagi conjecture \cite{Ryu:2006bv,Ryu:2006ef} that connects entanglement entropy of the boundary theory to the area of a minimal surface in the bulk. A disadvantage of these methods is that they do not provide the spectrum of the reduced density matrix, which contains the whole entanglement information. Actually, these methods can provide a family of extensions of the entanglement entropy, the so-called \Renyi entanglement entropies. In principle, the spectrum of the reduced density matrix can be recovered from the whole family, a difficult task in practice.

There is a more direct, brute-force method to calculate entanglement entropy in field theory. It uses textbook quantum mechanics of the simple harmonic oscillator, but requires part of the calculation to be done numerically. It relies on the discretization of the system to a harmonic system with a countable number of degrees of freedom, and it is applicable in practice when the state of the overall system is Gaussian. On the other hand, this method has the advantage that it provides the full reduced density matrix and its spectrum as an intermediate result. It was first employed by Srednicki \cite{Srednicki:1993im} for the derivation of the famous area-law property of entanglement entropy of free massless scalar field theory in flat space at its ground state. The leading correction to the area-law term is a logarithmic term with a universal coefficient. The entanglement entropy has an expansion of the form
\begin{equation}\label{eq:SEE_expansion_flat}
	S^{(3\textrm{d})}_{\textrm{flat}} = d_2 \frac{R^2}{\epsilon^2} + d_1 \ln \frac{R}{\epsilon} + d_0+\dots ,
\end{equation}
where $\epsilon$ is the UV cutoff length. (The scale factor is trivially $a=1$, and the physical cutoff coincides with the comoving one for a flat background.) The numerical determination of $d_1$ is very difficult because of the dominance of the area-law term. Its value has been calculated to be $d_1 = - {1}/{90}$ \cite{Lohmayer:2009sq}. Theoretical arguments based on holographic duality agree with this value \cite{Solodukhin:2008dh,Casini:2009sr}. The dots represent subleading terms of the form $\left(\epsilon/R\right)^{2i}$, where $i \in \mathbb{N}^{*}$, which are negligible in the continuum limit. This method and its extensions provide the basis for the calculations performed in this work.

We also quote the result for $(1+1)$-dimensional conformal field theories, which will also be relevant for our calculation:
\begin{equation}\label{eq:SEE_expansion_flat_1d}
	S^{(1\textrm{d})}_{\textrm{flat}}= \frac{c}{6} \ln \frac{r}{\epsilon} + d+\dots ,
\end{equation}
where $c$ is the central charge \cite{Holzhey:1994we,Korepin:2004zz,Calabrese:2004eu}. We have assumed Dirichlet boundary conditions for the overall system, whose length has been taken to be infinite.

\subsection{Entanglement entropy in flat space}
\label{subsec:review_Srednicki}

Let us consider a massless scalar field in flat space,
\begin{equation}\label{eq:action_flat}
\mathcal{S} = \frac{1}{2} \int dt \, d^3 \mathbf{x} \, \left(\dot{\phi}^2 - \left( \mathbf{\nabla} \phi \right)^2 \right).
\end{equation}
Since we are interested in spherical entangling surfaces, symmetry suggests the  expansion of the field $\varphi \left( \mathbf{x} \right)$ and its conjugate momentum $\pi \left( \mathbf{x} \right)$ in spherical harmonic moments:
\begin{align}
\phi_{\ell m} \left( r \right) &= r \int {d\Omega \, Y_{\ell m} \left( {\theta ,\varphi} \right) \phi \left( \mathbf{x} \right)} , \\
\pi_{\ell m} \left( r \right) &= r \int {d\Omega \, Y_{\ell m} \left( {\theta ,\varphi} \right) \pi \left( \mathbf{x} \right)} ,\label{eq:moments}
\end{align}
where $r$, $\theta$ and $\varphi$ are the usual three-dimensional spherical coordinates and $Y_{\ell m} \left( \theta ,\varphi \right)$ the real spherical harmonics.

In order to obtain a quantum mechanical system with countable degrees of freedom, the radial coordinate has to be discretized. This is done via the introduction of a lattice of concentric spherical shells with radii $r_j = j \epsilon$, where $j = 1 , 2 , \ldots , N$. The lattice introduces an UV cutoff energy scale equal to $1 / \epsilon$, while the size of the lattice $L = N \epsilon$ sets an IR cutoff energy scale equal to $1 / L$. We define the discretized degrees of freedom as
\begin{equation}
\phi_{\ell m} \left( {j \epsilon} \right) \to \phi_{\ell m,j} , \quad
\pi_{\ell m} \left( {j \epsilon} \right) \to \frac{\pi_{\ell m,j}}{\epsilon}, 
\end{equation}
so that they are canonically commuting. Then, the Hamiltonian reads
\begin{multline}
H = \frac{1}{2 \epsilon} \sum\limits_{\ell,m} \sum\limits_{j = 1}^N \Bigg[ \pi_{\ell m,j}^2 +  \frac{\ell\left( {\ell + 1} \right)}{j^2}   \phi_{\ell m,j}^2\\+ {\left( {j + \frac{1}{2}} \right)}^2 {\left( {\frac{\phi_{\ell m,j + 1}}{j + 1} - \frac{\phi_{\ell m,j}}{j}} \right)}^2  \Bigg] .
\label{eq:Hamiltonian_discretized_flat}
\end{multline}
The important property of the above Hamiltonian is the fact that the various angular-momentum sectors, each identified by a pair $\left\{ \ell , m \right\}$, do not interact with each other. Furthermore, the index $m$ does not appear at all in the dynamics. Therefore, the problem has been split into a countably infinite set of problems, one for each value of $\ell$. Assuming that the radius of the entangling surface equals 
\begin{equation}
R = \left( n + \frac{1}{2} \right) \epsilon\equiv n_R \epsilon,
\end{equation}
the entanglement entropy reads
\begin{equation}
S\left( n , N \right) = \sum_{\ell=0}^{\infty} \left( 2 \ell + 1 \right) S_\ell \left( n , N \right) ,
\label{eq:SEE_ell_sum}
\end{equation}
where $S_\ell$ is the bipartite entanglement entropy for a single $\ell$-sector, whose degrees of freedom have been split into the subsystem $A$, containing $\phi_{lm,j}$ with $j \leq n$, and its complementary subsystem $A^C$.

When the overall system lies at its ground state, the state of each $\ell$-sector reads
\begin{equation}
\Psi\left( \mathbf{x} \right) \sim \exp \left( - \frac{1}{2}\mathbf{x}^T \Omega \mathbf{x} \right) ,
\end{equation}
where $\Omega$ is the positive square root of the coupling matrix $K$ of the corresponding $\ell$-sector Hamiltonian. The latter can be directly read from equation \eqref{eq:Hamiltonian_discretized_flat}:
\begin{multline}
K_{ij} = \left( 2 +  \frac{\ell \left( \ell + 1 \right) + \frac{1}{2}}{i^2} - \frac{1}{4} \delta_{i,1} \right) \delta_{i , j} \\  - \frac{\left( i + \frac{1}{2} \right)^2}{i \left( i + 1 \right)} \delta_{i + 1 , j} - \frac{\left( j + \frac{1}{2} \right)^2}{j \left( j + 1 \right)} \delta_{i , j + 1}.
\label{eq:Coupling_Matrix}
\end{multline}
For the $(1+1)$-dimensional field theory, the coupling matrix reads
\begin{equation}
K^{(1\textrm{d})}_{ij} = 2\delta_{i , j} - \delta_{i + 1 , j} - \delta_{i , j + 1}.
\label{eq:Coupling_Matrix_1d}
\end{equation}
This is different than the expression resulting from \eqref{eq:Coupling_Matrix} for $\ell=0$, because it corresponds to a slightly different discretization scheme. Actually, their difference amounts only to a boundary term. Therefore, the result for the entropy in the continuum limit is the same for both schemes, a fact that we have verified numerically.

We introduce the block-form notation
\begin{equation}
\Omega = \left( \begin{array}{cc} \Omega_{A} & \Omega_{B} \\ \Omega_B^T & \Omega_C \end{array} \right) , \quad \mathbf{x} = \left( \begin{array}{c} \mathbf{x}_A \\ \mathbf{x}_C \end{array} \right) ,
\label{eq:blocks}
\end{equation}
where the vector $\mathbf{x}_A$ spans the subsystem $A$, and $\mathbf{x}_C$ its complement $A^c$. 
It is a matter of simple algebra to show that the reduced density matrix reads
\begin{multline}
\rho_A \left( \mathbf{x}_A ; \mathbf{x}_A^\prime \right) \sim \exp \bigg[ - \frac{1}{2} \left( \mathbf{x}_A^T \gamma \mathbf{x}_A  + \mathbf{x}_A^{\prime T} \gamma \mathbf{x}^\prime_A \right)  \\ + \mathbf{x}_A^{\prime T} \beta \mathbf{x}_A \bigg] ,\label{eq:reduced_density_matrix}
\end{multline}
where
\begin{equation}
\gamma = \Omega_A - \frac{1}{2} \Omega_B^T \Omega_C^{-1} \Omega_B , \quad 
\beta = \frac{1}{2} \Omega_B^T \Omega_C^{-1} \Omega_B .
\label{eq:gamma_beta}
\end{equation}

The spectrum of the reduced density matrix \eqref{eq:reduced_density_matrix} can be found analytically. It reads
\begin{equation}
\lambda_{\left\{ m_1 , m_2 , \ldots , m_n \right\}} = \prod_{i = 1}^n \left( 1 - \xi_i \right) \xi_i^{m_i} ,
\label{eq:spectrum_flat}
\end{equation}
where the indices $m_1 , m_2, \ldots , m_n$ are non-negative integers, $\xi_i = \frac{\beta_i}{1 + \sqrt{1 - \beta_i^2}}$ and $\beta_i$ are the eigenvalues of the matrix $\tilde{\beta} = \gamma^{-1} \beta$. It follows that the contribution of a single $\ell$-sector to the entanglement entropy reads
\begin{equation}
S_{\ell} = - \sum_{i = 1}^n \left( \ln \left( 1 - \xi_i \right) + \frac{\xi_i}{1 - \xi_i} \ln \xi_i \right) .
\label{eq:SEE}
\end{equation}

\subsection{Entanglement entropy in de Sitter space}
\label{subsec:Srednicki_dS}
We turn our attention to the case of a massless scalar field in a FRW background,
\begin{equation}
ds^2 = a^2 \left( \tau \right) \left( d\tau^2 - dr^2 - r^2 d\Omega^2 \right) ,
\label{eq:dsmetric}
\end{equation}
where $\tau$ is the conformal time. The redefinition of the field as $\phi \left( \tau , \mathbf{x} \right) = f \left( \tau , \mathbf{x} \right) / a \left( \tau \right)$, allows the expression of the action as
\begin{equation}
\mathcal{S} = \frac{1}{2} \int d\tau \, d^3 \mathbf{x} \, \left( \dot{f}^2 - \left( \nabla f \right)^2 + \frac{\ddot{a}}{a} f^2 \right),
\label{eq:action_dS}
\end{equation}
i.e. in a form where the kinetic term is canonical. The dot denotes differentiation with respect to the conformal time $\tau$.

The next steps in the calculation of the entanglement entropy are identical to the flat-space case; the only difference is the presence of the time-dependent mass term $-(\ddot{a}/a) f^2$. We expand the field in its spherical moments. We then introduce a lattice of spherical shells and arrive at a discretized Hamiltonian that reads
\begin{multline}
H = \frac{1}{2 \epsilon} \sum\limits_{\ell,m} \sum\limits_{j = 1}^N \Bigg[ \pi_{\ell m,j}^2+ \left( \frac{\ell \left( {\ell + 1} \right)}{j^2} - \epsilon^2 \frac{\ddot{a}}{a} \right) f_{\ell m,j}^2 \\
+ {\left( {j + \frac{1}{2}} \right)}^2 {\left( {\frac{f_{\ell m,j + 1}}{j + 1} - \frac{f_{\ell m,j}}{j}} \right)}^2 \Bigg] .
\label{eq:Hamiltonian_discretized_dS}
\end{multline}
The lattice spacing $\epsilon$, which acts as an UV cutoff, is now a comoving scale. As in the flat-space case, the discretized Hamiltonian is split into angular-momentum sectors that do not interact with each other. We focus on the case of de Sitter space, i.e.
\begin{equation}
a \left( \tau \right) = - \frac{1}{H \tau} , \quad \frac{\ddot{a}}{a} = \frac{2}{\tau^2}.
\label{eq:scale_factor}
\end{equation}

In the case of a flat background, the Hamiltonian \eqref{eq:Hamiltonian_discretized_flat} is time-independent and can be written as the direct sum of simple harmonic oscillators via an orthogonal transformation of the coordinates; each one of those corresponds to a single normal mode of the system. The ground state of the overall system is obviously the tensor product of the ground states of all these normal-mode oscillators. In the case of a dS background, the same orthogonal transformation allows the expression of the Hamiltonian \eqref{eq:Hamiltonian_discretized_dS} as a direct sum of Hamiltonians that contain a single degree of freedom. However, these are not simple harmonic-oscillator Hamiltonians, due to the existence of the time-dependent mass term in equation \eqref{eq:action_dS}. If the Hamiltonian of a single normal mode in the case of flat space is
\begin{equation}
H^{\textrm{flat}}_{\textrm{normal mode}} = \frac{1}{2} p^2 + \frac{1}{2} \omega_0^2 x^2 ,
\label{eq:Schrodinger_time_independent}
\end{equation}
then, in the case of a dS background, the corresponding Hamiltonian will be
\begin{equation}
H^{\textrm{dS}}_{\textrm{normal mode}} = \frac{1}{2} p^2 + \frac{1}{2} \left( \omega_0^2 - \frac{2 \epsilon^2}{\tau^2} \right) x^2 ,
\label{eq:Schrodinger_time_dependent}
\end{equation}
i.e. it corresponds to a deformation of the single harmonic oscillator with a time-dependent eigenfrequency. Notice that the appropriate dimensionless time variable for the evolution of the system is $\tau/\epsilon$, while the spatial directions are parameterized in terms of the integers $\ell$, $m$ and $j$. In the following we set $\epsilon=1$, with the assumption that all dimensionful quantities are measured in units of $\epsilon$. The cutoff can be reinstated explicitly through dimensional analysis when needed, as will be done at certain points in the paper.

The Hamiltonian \eqref{eq:Schrodinger_time_dependent} does not possess energy eigenstates. We assume that each one of these one-degree-of-freedom systems is described by a time-dependent wave function that reduces to the ground-state wave function of the simple harmonic oscillator as $\tau \to - \infty$. This corresponds to the choice of the Bunch-Davis vacuum. The wave function can be systematically constructed. First, one solves the Ermakov equation
\begin{equation}
\ddot{b} \left( \tau \right) + \left( \omega_0^2 - \frac{2}{\tau^2} \right) b \left( \tau \right) = \frac{\omega^2_0}{b^3 \left( \tau \right)},
\label{eq:Ermakov}
\end{equation} 
with an appropriate initial condition, which in our case is $b(\tau)\to 1$ for $\tau \to -\infty$. Then, any solution $F^0 \left( \tau , x \right)$ of the Schr\"odinger equation with the simple harmonic oscillator Hamiltonian \eqref{eq:Schrodinger_time_independent} can be upgraded to a solution $F \left( \tau , x \right)$ of the time-dependent Schr\"odinger equation with the Hamiltonian \eqref{eq:Schrodinger_time_dependent} as follows:
\begin{multline}
F \left( \tau , x \right) = \frac{1}{\sqrt{b \left( \tau \right)}} \, \exp \left({\frac{i}{2} \frac{\dot{b} \left( \tau \right)}{b \left( \tau \right)} x^2} \right) \\ \times F^0 \left( \int \frac{d\tau}{b^2 \left( \tau \right)} , \frac{x}{b \left( \tau \right)} \right) .
\label{eq:sol_schrod}
\end{multline}

It is straightforward to verify that the solution of the Ermakov equation is
\begin{equation}
b^2 \left( \tau \right) = 1 + \frac{1}{\omega_0^2\tau^2} ,
\end{equation}
which yields
\begin{equation}\label{groundstatee}
F \left( \tau , x \right) \propto \exp \left[ - \frac{1}{2} \omega_0 \left( 1 - \frac{1}{\omega_0 \tau}\frac{1} {i + \omega_0 \tau} \right) x^2 \right] .
\end{equation}
This wave function reduces to the ground state of the simple harmonic oscillator \eqref{eq:Schrodinger_time_independent} as $\tau \to - \infty$. Furthermore, it is Gaussian at all times. It follows that for each $\ell$-sector the wave function reads
\begin{equation}
\Psi \left( \mathbf{x} \right) \propto \exp \left( - \frac{1}{2} x^T W x \right),
\label{eq:wavefunction_dS}
\end{equation}
where the components of the vector $x$ are $x_i = \phi_{lm,i}$. The matrix $W$ is complex symmetric and is given by
\begin{equation}
W = \Omega \left[ I - \left( I - \frac{i}{ \tau}\Omega^{-1} \right) \left(I + \Omega^2 \tau^2\right)^{-1}  \right] ,
\label{eq:matrix_W}
\end{equation}
where $\Omega$ is the positive square root of the coupling matrix \eqref{eq:Coupling_Matrix}, i.e. the coupling matrix in the case of flat space. 

The wave function is Gaussian at all times, allowing for a simple calculation of the reduced density matrix. However, since the matrix \eqref{eq:matrix_W} that appears in the exponent of the wave function is complex, the determination of the entanglement spectrum is much more involved. As expected, the matrices $\gamma$ and $\beta$ \eqref{eq:gamma_beta} that appear in the reduced density matrix become complex too. It can be shown that the spectrum is still of the form \eqref{eq:spectrum_flat} where $\xi_i$ are the eigenvalues of some matrix that depends on $\gamma$ and $\beta$ \cite{Katsinis:2023hqn}. However, the specification of this matrix requires the solution of a quadratic matrix equation, and there is an additional difficulty in determining which solution of this equation is the appropriate one. The problem can be bypassed by solving an eigenvalue problem with double the dimension of the original one. For the details we refer the reader to \cite{Katsinis:2023hqn}.

The approach outlined above is equivalent to the calculation of the entanglement spectrum via the covariance matrix method, as shown in the appendix E of \cite{Katsinis:2023hqn}. We consider the $2 n \times 2 n$ matrix $\mathcal{M}$ given by
\begin{equation}
\mathcal{M} = 2i J \,\mathrm{Re}  M,
\end{equation}
where the covariance matrix  $M$ and the matrix $J$ are defined as
\begin{equation}
M = \begin{pmatrix}
\left\langle x_i x_j \right\rangle & \left\langle x_i \pi_j\right\rangle \\
\left\langle x_i \pi_j \right\rangle^T & \left\langle \pi_i \pi_j \right\rangle
\end{pmatrix} , \quad J = \begin{pmatrix}
0 & I \\
-I & 0
\end{pmatrix} .
\end{equation} 
Employing the wave function \eqref{eq:wavefunction_dS} we find 
\begin{small}
\begin{multline}
\mathcal{M} =\\ i\begin{pmatrix}
- \im W   \left(\re W \right)^{-1} & \re W + \im W \left( \re W \right)^{-1} \im W \\
- \left( \re W \right)^{-1} & \left(\re W \right)^{-1} \im W 
\end{pmatrix} .
\end{multline}
\end{small}
Substituting \eqref{eq:matrix_W} and restricting to subsystem $A$, we obtain the corresponding matrix $\mathcal{M}$ that reads
\begin{small}
\begin{multline}\label{eq:matrix_M_cal}
\mathcal{M} = \\ i\begin{pmatrix}
\frac{\left( \Omega^{-3} \right)_A}{\tau^3} &  \left(\Omega\right)_A - \frac{\left( \Omega^{-1} \right)_A}{\tau^2}+\frac{\left( \Omega^{-3} \right)_A}{\tau^4}  \\
- \left( \Omega^{-1} \right)_A - \frac{\left( \Omega^{-3} \right)_A}{\tau^2} &\quad-\frac{\left( \Omega^{-3} \right)_A}{\tau^3} 
\end{pmatrix} .
\end{multline}
\end{small}
The eigenvalues of this matrix come in pairs $\pm \lambda_i$ and they all  satisfy $\left| \lambda_i \right| \geq 1$. The entanglement entropy is given in terms of the eigenvalues of the matrix $\mathcal{M}$ as
\begin{equation}\label{eq:SEE_of_M_cal}
S = \sum_{i=0}^{n} \left( \frac{\lambda_i + 1}{2} \ln \frac{\lambda_i + 1}{2} - \frac{\lambda_i - 1}{2} \ln \frac{\lambda_i - 1}{2} \right),
\end{equation}
where the sum is performed over the positive eigenvalues only.

In an obvious manner, when the matrix $W$ is real the diagonal blocks of the covariance matrix $\mathcal{M}$ vanish and the eigenvalue problem becomes equivalent to a problem of half the dimension. It can be shown that this is equivalent to the problem solved in section \ref{subsec:review_Srednicki}.

\subsection{Perturbation theory}
\label{subsec:expansion}

The structure of the matrix $\mathcal{M}$ given in \eqref{eq:matrix_M_cal} implies that its eigenvalues and consequently the entanglement entropy have a non-trivial dependence on $\tau$. This is the reason for considering the general parameterization of equation \eqref{eq:SEE_expansion} for the entropy. Obtaining a closed analytical expression that is valid for any time seems very difficult.  The numerical determination of the various terms would be greatly facilitated by some analytical intuition. One possible resolution of this issue is the use of perturbation theory in order to constrain the dependence of the entanglement entropy on the various parameters in certain limits. 

Naturally there are two expansions, one for early times and a complementary one for late times. In the late-time limit $|\tau|/\epsilon=(1/H)/(a\epsilon) =(1/H)/\epsilon_p\to 0$, the discretization of the theory is problematic, as the physical UV cutoff length becomes longer than the Hubble radius. Mathematically this limit is well defined, but the physical significance of its results is questionable. However, an expansion around $\tau=0$ can capture terms proportional to $\ln|\tau|$ that are not visible in an expansion around $\tau=-\infty$. In the following we develop the more physical $1/\tau$ expansion in order to compare with the numerical results of the following sections.

Let us consider the indicative, exactly solvable case in which the reduced system has a single degree of freedom. In this case, the matrix $\mathcal{M}$ reads
\begin{small}
\begin{multline}
\mathcal{M}=\\ i\begin{pmatrix}
\frac{\left(\omega^{-3}\right)_A}{\tau^3} & \big(\omega\big)_A-\frac{\left(\omega^{-1}\right)_A}{\tau^2}+\frac{\left(\omega^{-3}\right)_A}{\tau^4}\\
-\left(\omega^{-1}\right)_A -\frac{\left(\omega^{-3}\right)_A}{\tau^2} &  \quad-\frac{\left(\omega^{-3}\right)_A}{\tau^3} 
\end{pmatrix}
\label{eq:app_M_tilde_unrotated}
\end{multline}
\end{small}
and its eigenvalues are
\begin{equation}\label{eq:eigs_2_osc}
\lambda=\pm \sqrt{\left(\omega^{-1}\right)_A\left(\omega\right)_A+\frac{\left(\omega^{-3}\right)_A\big(\omega\big)_A-\left(\omega^{-1}\right)_A^2}{\tau^2}}.
\end{equation}
The form of these eigenvalues indicates that  their early-time expansion starts from $\tau^0$ and has only even powers of $1/\tau$. The fact that the expansion of the eigenvalues of the full problem is of this form becomes apparent by considering the eigenvalue problem of the square of the matrix $\mathcal{M}$. We perform an appropriate similarity transformation\footnote{The appropriate matrix $O$ for the similarity transformation
	is $O=\begin{pmatrix}
		\left(-\tau\right)^{-1/2} & 0 \\
		0 & \left(-\tau\right)^{1/2}
	\end{pmatrix}$.} to the matrix $\mathcal{M}^2$, so that only even powers of time appear, yielding the matrix $\tilde{\mathcal{M}}^2$, which reads
\begin{widetext}
\begin{equation}\label{eq:M_cal_sq}
\tilde{\mathcal{M}}^2=\begin{pmatrix} \frac{\mathcal{M}^{(0)T}}{\tau^0}+\frac{\mathcal{M}^{(-2)T}}{\tau^2}+\frac{\mathcal{M}^{(-4)}}{\tau^4} & \frac{\mathcal{M}^{(-2)}-\mathcal{M}^{(-2)T}}{\tau^2}-\frac{\mathcal{M}^{(-4)}}{\tau^4}\\ 
\frac{\mathcal{M}^{(-4)}}{\tau^4} & \frac{\mathcal{M}^{(0)}}{\tau^0}+\frac{\mathcal{M}^{(-2)}}{\tau^2}-\frac{\mathcal{M}^{(-4)}}{\tau^4}
\end{pmatrix},
\end{equation}
\end{widetext}
where
\begin{align}
\mathcal{M}^{(0)}\thickspace &= \left(\Omega^{-1}\right)_A \big(\Omega\big)_A,\\
\mathcal{M}^{(-2)} &= \left(\Omega^{-3}\right)_A\big(\Omega\big)_A-\left(\Omega^{-1}\right)_{A}^2,\\
\mathcal{M}^{(-4)} &= \left(\Omega^{-3}\right)_A \left(\Omega^{-1}\right)_A-\left(\Omega^{-1}\right)_{A}\left(\Omega^{-3}\right)_{A}.
\end{align}

Recall that the eigenvalues of $\mathcal{M}$ come in pairs of the form $(\lambda_i,-\lambda_i)$. As a result, all eigenvalues of $\tilde{\mathcal{M}}^2$ have multiplicity equal to two. Finally, as a trivial consistency check, we observe that, when considering as subsystem $A$ the overall system, all $A$-blocks of the powers of $\Omega$ appearing in the above relations become the whole matrices and therefore they commute. This makes the matrix $\tilde{\mathcal{M}}^2$ equal to the identity matrix $I_{2N}$, which obviously implies that the entanglement entropy vanishes.

We next develop the perturbative expansion for early times. The eigenvalue problem we are interested in solving reads
\begin{equation}
	\tilde{\mathcal{M}}^2 V_i=\Lambda_i V_i,
\end{equation}
where $\Lambda_i=\lambda_i^2$. We express the eigenvectors $V_i$ as ``doublets'' of $n$-dimensional vectors. In the flat-space limit the eigenvalue problem assumes the form
\begin{equation}
\begin{pmatrix}
		\mathcal{M}^{(0)T} & 0\\ 0 & \mathcal{M}^{(0)}
	\end{pmatrix}\begin{pmatrix}
		v_i^{(0)}\\
		w_i^{(0)}
	\end{pmatrix}=\Lambda_i^{(0)}\begin{pmatrix}
		v_i^{(0)}\\
		w_i^{(0)}
	\end{pmatrix}.
\end{equation}
This means that $w^{(0)}_i$ and $v^{(0)}_i$ are the right and left eigenvectors of the matrix $\mathcal{M}^{(0)}$, respectively. This fact makes apparent how the double degeneracy of the eigenvalues works in the flat-space limit. The eigenvectors can be divided to two classes, namely
\begin{equation}
	V_i^{(0)}=\begin{pmatrix}
		v^{(0)}_i\\
		0
	\end{pmatrix} ,\quad V_i^{\prime(0)}=\begin{pmatrix}
		0\\
		w^{(0)}_i
	\end{pmatrix},
\end{equation}
which correspond to the eigenvalue $\Lambda_i^{(0)}$. This degeneracy in not lifted at any order of perturbation theory, because it is a consequence of the fact that the eigenvalues of $\mathcal{M}$ come in pairs of the from $\pm \lambda_i$ for any $\tau$. Therefore, regarding the specification of the eigenvalues, it suffices to consider only half of the eigenvectors.

The matrix $\mathcal{M}^{(0)}$ can be made symmetric via a similarity transformation. It follows that when the spectrum of $\mathcal{M}^{(0)}$ is non-degenerate\footnote{When the subsystem $A$ is larger than its complementary, there are degenerate eigenvalues that are equal to one. In order to apply perturbation theory one should introduce a basis in this subspace. However, this would not generate additional complexity in perturbation theory, as these eigenvalues do not receive any corrections.}, the eigenvectors $w^{(0)}_i$ are linearly independent. This is the case for $v^{(0)}_i$ too. Being the left and right eigenvectors of  $\mathcal{M}^{(0)}$, they obey
\begin{equation}
	v^{(0)T}_i w^{(0)}_j = N_i\delta_{ij},
\end{equation}
where $N_i$ is a normalization factor that we do not fix explicitly.

The form of the matrix \eqref{eq:M_cal_sq} suggests that the eigenvalues and the eigenvectors have expansions of the form
\begin{equation}
V_i=\begin{pmatrix}
		v_i\\
		u_i
	\end{pmatrix},\quad v_i = \sum_{k=0}^\infty \frac{v^{(2k)}_i}{\tau^{2k}},\quad	u_i = \sum_{k=1}^\infty \frac{u^{(2k)}_i}{\tau^{2k}}
\end{equation}
and
\begin{equation}
\Lambda_i = \sum_{k=0}^\infty \frac{\Lambda^{(2k)}_{i}}{\tau^{2k}},
\end{equation}
where $i=1,\dots, n$. The perturbative expansion is built around the flat-space result,
\begin{equation}\label{eq:expansioneig_early}
	\lambda_i=\lambda_i^{(0)}+\frac{1}{\tau^2}\frac{\Lambda^{(2)}_i}{2\lambda_i^{(0)}}+\dots,
\end{equation}
where $\lambda_i^{(0)}=\sqrt{\Lambda^{(0)}_i}$ are the eigenvalues of the flat-space problem and $\Lambda^{(2)}_{i}$ are the leading de Sitter corrections.

At first order in perturbation theory we obtain
\begin{align}
	\mathcal{M}^{(0)T}v^{(2)}_i+\mathcal{M}^{(-2)T}v^{(0)}_i &= \Lambda^{(2)}_{i} v^{(0)}_i +\Lambda^{(0)}_{i} v^{(2)}_i , \\
	\mathcal{M}^{(0)}u^{(2)}_i&= \Lambda^{(0)}_{i} u^{(2)}_i .
\end{align}
Projecting the first equation on $w^{(0)}_i$, we obtain
\begin{equation}
\begin{split}
\Lambda^{(2)}_{i}&=\frac{v^{(0)T}_i\mathcal{M}^{(-2)}w^{(0)}_i}{v^{(0)T}_i w^{(0)}_i}\\
&=\frac{v^{(0)T}_i\left[\left(\Omega^{-3}\right)_A\big(\Omega\big)_A-\left(\Omega^{-1}\right)_{A}^2\right]w^{(0)}_i}{v^{(0)T}_i w^{(0)}_i} .
\end{split}
	\label{lambdamtwo}
\end{equation}

Substituting the expansion \eqref{eq:expansioneig_early} in \eqref{eq:SEE_of_M_cal}, the entanglement entropy of each angular-momentum sector assumes the form
\begin{equation}\label{eq:SEE_expansion_early}
	S_{\ell,\textrm{dS}} = S_{\ell,\textrm{flat}}+ \frac{1}{\tau^2}\sum_{i=0}^{n} \frac{\Lambda^{(2)}_{\ell,i}}{2\lambda_{\ell,i}^{(0)}}\textrm{arccoth}\,\lambda_{\ell,i}^{(0)}+\dots
\end{equation}

There are several lessons that can be drawn from perturbation theory, which will be relevant for the numerical analysis. At sufficiently early times, the leading contribution to the entanglement entropy is that for flat space. The first correction is of order $1/\tau^2$, which  corresponds to a factor $\sim H^2$ when the entropy is expressed in terms of physical lengths. No corrections $\sim \ln|\tau|$ appear in this limit, as the expansion is performed around a regular point. It is expected that the corrections arising through the cosmological expansion will introduce at most subleading UV divergences. This means that the appearance of $H^2$ must be accompanied by another dimensionful quantity, such as the radius of the entangling surface $R_p$ in the combination $H^2 R^2_p$. It is then apparent that the convergence of the perturbative expansion limits us to subhorizon subsystems with $R_p  <  1/H$.

There is also a rather surprising conclusion that can be reached on purely analytical grounds. The matrix $\Omega$ that underlies perturbation theory is the positive square root of the coupling matrix $K$, which assumes the form \eqref{eq:Coupling_Matrix} for the $(3+1)$-dimensional theory and \eqref{eq:Coupling_Matrix_1d} for the $(1+1)$-dimensional theory. The $\ell=0$ sector of the $(3+1)$-dimensional theory is essentially equivalent to the $(1+1)$-dimensional theory. The difference between \eqref{eq:Coupling_Matrix} for $\ell=0$ and \eqref{eq:Coupling_Matrix_1d} is a difference in the discretization scheme and is irrelevant in the continuum limit. For this reason, we can  draw conclusions for the role of the $\ell=0$ sector by considering the  simpler coupling structure of \eqref{eq:Coupling_Matrix_1d}.

The eigenvalue problem
\begin{equation}
	K_{ij} v_j=\lambda v_i
\end{equation}
in the continuum limit becomes the differential equation
\begin{equation}
	-f^{\prime\prime}(x)= \lambda f(x),\quad f(0)=f(L)=0,
\end{equation}
where $L$ stands for the size of the \emph{overall system}.
As a result, we obtain
\begin{equation}
	K(x,x^\prime)=\frac{2}{L}\sum_{k=1}^\infty
	\frac{k^2\pi^2}{L^2}\sin\frac{k\pi x}{L}\sin\frac{k\pi x^\prime}{L}.
\end{equation}
The powers of the matrix $\Omega$ in the continuum limit
become the kernels
\begin{equation}
	\Omega^m (x,x^\prime)=\frac{2}{L}\sum_{k=1}^\infty
	\frac{k^m\pi^m}{L^m}\sin\frac{k\pi x}{L}\sin\frac{k\pi x^\prime}{L}.
\end{equation}
In particular, for $\Omega^{-3}$ we obtain
\begin{multline}
	\Omega^{-3}(x,x^\prime)=\frac{L^2}{2\pi^3}\Big[\mathrm{Li}_3\left(e^{-i\frac{\pi}{L}\left(x-x^\prime\right)}\right)\\ -\mathrm{Li}_3\left(e^{-i\frac{\pi}{L}\left(x+x^\prime\right)}\right)+\textrm{c.c.}\Big],
\end{multline}
where $\mathrm{Li}_3(x)$ is the polylogarithm of order $3$. For $L\gg x$ and
$L\gg x^\prime$ the kernel assumes the form
\begin{multline}
	\Omega^{-3}(x,x^\prime)=\frac{1}{2\pi}\bigg[6x x^\prime+\left(x-x^\prime\right)^2\ln \frac{\pi\left\vert x-x^\prime\right\vert}{L}\\ -\left(x+x^\prime\right)^2\ln \frac{\pi\left\vert x+x^\prime\right\vert}{L}\bigg].
\end{multline}
On the other hand, neither $\Omega(x,x^\prime)$ nor $\Omega^{-1}(x,x^\prime)$
depends on $L$ in this limit \cite{Katsinis:2024gef}.

It is then apparent that a contribution involving $\ln L$ is expected in the entanglement entropy. The leading term in $H$ turns out to be $\sim H^2 R_p^2 \ln(L_p/\epsilon_p)$. The proportionality constant can be derived analytically, through a calculation that will be presented elsewhere \cite{Boutivas:2024sat}. The above discussion justifies the inclusion of the last term in \eqref{eq:SEE_expansion}. It must be emphasized that the presence of a term involving $\ln(L_p/\epsilon_p)$ is not limited to the region of validity of perturbation theory, as the matrix $\Omega^{-3}$ appears in the exact expression for the entropy. In the context of the perturbative scheme that we employed this term cannot become dominant, as sending $L\to \infty$ in order to increase the magnitude of the  $\ln(L_p/\epsilon_p)$ term invalidates the expansion. The regime of validity of our expansion is not restricted to just subhorizon entangling surfaces, but actually to ones that obey $H^2 R_p^2 \ln(L_p/\epsilon_p)<1$. However, this logarithmic term represents the leading correction induced by the expansion of the background.

\section{Numerical analysis}
\label{sec:numerical_analysis}
\subsection{State of the art}
\label{subsec:review_numerical_flat}

The method described in section \ref{subsec:review_Srednicki} was used in \cite{Srednicki:1993im} in order to compute the leading area-law term  of entanglement entropy in massless free scalar field theory at its ground state. The computational requirements of this task are trivial with nowadays computing power. A more difficult task is the calculation of the subleading logarithmic term in equation \eqref{eq:SEE_expansion_flat} \cite{Lohmayer:2009sq}. An obvious difficulty that arises is the dominance of the area-law term by several orders of magnitude. It follows that the calculation has to be performed with enough precision so that the logarithmic term is tractable. However, this is not the only technical difficulty that arises. Another important one has to do with effects due to the finite size of the overall system, which inevitably appear in numerical calculations. The expansion \eqref{eq:SEE_expansion_flat} is valid for a field theory defined on infinite flat space. The definition of the theory on a lattice of spherical shells implies that the entanglement entropy  depends on the IR cutoff. The existence of finite-size effects in a discretized theory can be made obvious considering that $S_\ell \left( N , N \right)$ must vanish. This cannot be the outcome of the two terms appearing in the expansion \eqref{eq:SEE_expansion_flat}; it is the outcome of their competition with the finite-size terms.

The accurate calculation of the logarithmic term in the infinite-size theory requires the subtraction of these finite-size effects. By varying the size of the overall system, while keeping the size of the subsystem fixed, one can check that the dominant finite-size correction to the entanglement entropy in a single angular-momentum sector behaves like $N^{-2\left(\ell +1\right)}$.

The finite-size effects are more suppressed in high angular-momentum sectors. This is a consequence of the fact that the angular momentum acts effectively as a position-dependent mass, localizing the form of the normal modes. As a result, the higher the angular momentum, the more local the nature of entanglement entropy becomes. Thus it is less sensitive to the size of the overall system. The required precision for this calculation is estimated by the authors of \cite{Lohmayer:2009sq} to be $10^{-8}$. As a consequence only a small subset of angular-momentum sectors requires the subtraction of these finite-size corrections in order to achieve the required precision. In particular, the authors mention that already for $\ell = 20$ and $n = 20$ the difference between $N_0 = 60$ and larger overall systems is of order $10^{-21}$.

After the subtraction of the finite-size effects, the calculation of the entanglement entropy requires the addition of the contributions of all angular-momentum sectors. However these are infinite, so in practice an angular momentum cutoff $\ell_{\textrm{max}}$ must be introduced. Through use of this cutoff the sum \eqref{eq:SEE_ell_sum} is split into a truncated sum $S_{\infty} \left( n ; \ell_{\textrm{max}} \right)$ and a remainder $Q_{\infty} \left( n ; \ell_{\textrm{max}} \right)$, as
\begin{equation}
S_{\infty} \left( n \right) = S_{\infty} \left( n ;\ell_{\textrm{max}} \right) + Q_{\infty}\left( n ;\ell_{\textrm{max}} \right) ,
\end{equation}
where
\begin{align}
S_{\infty} \left( n ; \ell_{\textrm{max}} \right) &= \sum_{\ell = 0}^{\ell_{\textrm{max}}} \left( 2 \ell + 1 \right) S_{\ell,\infty} \left( n \right) , \label{eq:SEE_sum_truncated} \\
Q_{\infty} \left( n ; \ell_{\textrm{max}} \right) &= \sum_{\ell = \ell_{\textrm{max}} + 1}^{\infty} \left( 2 \ell + 1 \right) S_{\ell,\infty} \left( n \right) . \label{eq:SEE_sum_reminder}
\end{align}
The subscript $\infty$ denotes the infinite-size limit. As we discussed, the modes of the high angular-momentum sectors are localized, which leads to a natural suppression of finite-size corrections. In a similar manner, high angular-momentum sectors have a limited contribution to the overall entanglement entropy. The effect of these sectors can be approximated by a hopping expansion \cite{Srednicki:1993im,Katsinis:2017qzh}. By setting a very large $\ell_{\textrm{max}}$, one can approximate the remainder $Q_{\infty}\left( n ;\ell_{\textrm{max}} \right)$ accurately enough using this expansion. The truncated sum $S_{\infty}\left( n ;\ell_{\textrm{max}} \right) $ must be calculated numerically.

This strategy implemented with the use of Wolfram's Mathematica, mostly using machine precision, results in the values \cite{Lohmayer:2009sq}
\begin{equation}
d_2 = 0.295431 , \quad d_1 = - 0.01109
\end{equation}
for the coefficients of equation \eqref{eq:SEE_expansion_flat}. The latter is consistent with the value $d_1 = - {1}/{90}$ that can be deduced via holographic arguments \cite{Solodukhin:2008dh,Casini:2009sr}.

\subsection{Refining the methodology}
\label{subsec:Our_Methodology}

In our previous work we studied the entanglement entropy for de Sitter space and a radiation-dominated universe \cite{Boutivas:2023ksg,Boutivas:2023mfg}. The time-dependent entanglement entropy $S_{\ell}\left(n,N,\tau\right)$ was calculated for sectors of fixed angular momentum $\ell$, and the total entropy through the sum
\begin{equation}\label{eq:SEE_sum_dS}
S\left(n,N,\tau\right)=\sum_{\ell=0}^{\infty}\left(2\ell+1\right) S_{\ell}\left(n,N,\tau\right).
\end{equation}
When the entanglement entropy is proportional to the total number of degrees of freedom of the subsystem, as was found in the case of a radiation-dominated universe, this sum is not necessarily convergent. However, for large angular momenta, $S_{\ell}\left(n,N,\tau\right)$ behaves as the similar quantity in flat space, implying that convergence is guaranteed. We can then adopt the original approach of Srednicki \cite{Srednicki:1993im}. 

For our calculation we follow the procedure described in section \ref{subsec:review_numerical_flat} with several modifications.  These include:
\begin{itemize}
\item{Extent of finite-size corrections:} We have to study the finite-size corrections for many more angular momenta. In flat space, the leading correction for each  sector scales as $N^{-2(\ell+1)}$. It is natural to assume that the correction depends on the ratio $n/N$. As a result when studying a system consisting of  $n<n_0$, we demand
\begin{equation}
\left(\frac{n_0}{N_0}\right)^{2(\ell+1)}\leq 10^{-35},
\label{precis}
\end{equation}
which is the numerical precision of our code. Selecting $n_0 = 50$ implies that for $N_0 = 60$ we have to subtract the  corrections for all sectors up to $\ell=220$. For $\ell$ larger than 220 the corrections are comparable with the numerical precision of the calculation, thus they are neglected. Equation \eqref{precis} is just an order of magnitude estimate. However, the contribution to the entanglement entropy for $\ell\sim 200$ is much more suppressed than that of the $\ell=0$ sector, so that this estimate is sufficient. We emphasize that we take into account angular-momentum sectors up to $\ell \sim 10^6$ 
in order to achieve the required accuracy of our calculation. 
\item{Precision of the finite-size corrections:} We aim at a precision of $10^{-20}$ for the entanglement entropy for given $n$ and $\tau$ after adding all angular momentum contributions and accounting for the finite-size corrections. This means that we have to include subleading corrections in the fit with respect to $N$. In \cite{Lohmayer:2009sq} the authors calculated the difference $S_{\ell}\left(n,N\right)-S_{\ell}\left(n,N_0\right)$ and plotted it as a function of $N^{-2(\ell+1)}$, revealing that their relation is linear. In the dS case we find it more convenient to use log-log plots, since the finite-size corrections have a more complicated behaviour. We calculate the finite differences and we show that
\begin{multline}
\ln\left[\frac{S_{\ell}\left(n,N+\Delta N,\tau\right)-S_{\ell}\left(n,N,\tau\right)}{\Delta N}\right] \\ \simeq -a \ln N +b,
\end{multline}
which implies that
\begin{multline}
S_{\ell}\left(n,N,\tau\right) \simeq S_{\ell,\infty}\left(n,\tau\right) \\+\begin{cases}
\frac{S_{\ell}^{(a-1)}\left(n,\tau\right)}{N^{a-1}}, & a\neq 1,\\
S_{\textrm{IR}}\left(n,\tau\right)\ln N ,& a = 1.
\end{cases}
\end{multline}
To identify the scaling of subleading corrections we calculate 
\begin{multline}
\ln\left[\frac{S_{\ell}\left(n,N+2\Delta N,\tau\right)-S_{\ell}\left(n,N+\Delta N,\tau\right)}{\left(N+\Delta N\right)^{-a}\left(\Delta N\right)^2}\right.\\ \left.-\frac{S_{\ell}\left(n,N+\Delta N,\tau\right)-S_{\ell}\left(n,N,\tau\right)}{N^{-a}\left(\Delta N\right)^2}\right] \\ \simeq -a^\prime \ln N +b^\prime,
\end{multline}
which implies that
\begin{multline}
S_{\ell}\left(n,N,\tau\right) \simeq S_{\ell,\infty}\left(n,\tau\right) \\+\begin{cases}
\frac{S_{\ell}^{(a-1)}\left(n,N,\tau\right)}{N^{a-1}}+\frac{S_{\ell}^{(a+a^\prime-2)}\left(n,N,\tau\right)}{N^{a+a^\prime-2}},& a\neq 1,\\
S_{\textrm{IR}}\left(n,\tau\right)\ln N +\frac{S_{\ell}^{(a^\prime -1)}\left(n,N,\tau\right)}{N^{a^\prime-1}},&a = 1.
\end{cases}
\end{multline}
With this strategy we can identify the scaling of even more subleading terms.
\item{Summation over the contributions of all angular-momentum sectors:}  Instead of estimating the terms of the remainder of the sum $Q_{\infty}\left( n; \ell_{\textrm{max}} \right)$, see \eqref{eq:SEE_sum_reminder}, using a hopping expansion, we study the behaviour of the truncated sum $S_{\infty}\left( n; \ell_{\textrm{max}} \right)$, see \eqref{eq:SEE_sum_truncated}, as a function of $\ell_{\textrm{max}}$. The $\ell_{\textrm{max}}\rightarrow\infty$ limit is obtained using a fit. By including several subleading terms we obtain a much more precise estimate for the infinite sum over all angular momenta.
\item{Fits for $\tau$ and $n$:} We work in the regime of large $\tau$. We identify the flat-space limit and compare with the known results. An increased number of subleading corrections is taken into account (see next section). Finally, we fit the data for $10\leq n \leq 50$, including terms with powers of $1/n^2$ for increased precision. In all fits we implement the strategy described above for the calculation of subleading corrections.
\end{itemize}

\subsection{Numerical analysis strategy}
\label{subsec:Summary_Numerical}

We performed high-precision numerical calculations of the entanglement entropy for the $(1+1)$- and $(3+1)$-dimensional massless scalar field theory in flat space, the $(1+1)$-dimensional toy model in de Sitter space also studied in \cite{Boutivas:2023ksg}, and the $(3+1)$-dimensional scalar theory in de Sitter space that was studied in \cite{Boutivas:2023mfg}. The $(1+1)$-dimensional toy model corresponds to the vanishing-angular-momentum sector of the $(3+1)$-dimensional  theory. For de Sitter space, this correspondence requires an appropriate non-minimal coupling to gravity to be included in this toy model \cite{Boutivas:2023ksg}. We used a custom C++ code enabling very fast numerical calculations while working with 128-bit precision. In appendix \ref{sec:code} we describe the basic features of the code. The calculations in flat space serve as test cases for the validation of our code and their results are used in order to compare with the extrapolation of the de Sitter calculations to the flat-space limit.

Our analysis consists of the following steps:
\begin{enumerate}
\item We determine and subtract the finite-size effects for each angular-momentum sector with $\ell\leq 220$. To this purpose, we analyze the data as functions of $N$ for fixed values of $n$ and $\tau$. For de Sitter space, the $\ell=0$ sector has a term growing like $\ln N$. After a careful analysis, we conclude that the appropriate form of the expansions is
\begin{equation}\label{eq:vac_3d_N_corrections}
S_{\ell,\textrm{flat}}(n,N)=S_{\ell,\textrm{flat},\infty}(n)+\sum_{k=0}^{k_{\textrm{max}}}\frac{S_{\ell,\textrm{flat}}^{(k)}(n)}{N^{(2\ell+2+k)}}
\end{equation}
and
\begin{multline}
\label{eq:dS_3d_N_corrections}
S_{\ell,\textrm{dS}}(n,\tau,N)=S_{\textrm{IR}}(n,\tau)\left(\ln N\right)\delta_{\ell,0}\\ +S_{\ell,\textrm{dS},\infty}(n,\tau)
+\sum_{\substack{k=0\\2\ell+k\neq0}}^{k_{\textrm{max}}}\frac{S_{\ell,\textrm{dS}}^{(k)}(n,\tau)}{N^{(2\ell+k)}}.
\end{multline}
For $\ell\geq 221$ the finite-size effects are considered negligible, implying that
\begin{align}
S_{\ell,\textrm{flat}}(n,N)&=S_{\ell,\textrm{flat},\infty}(n),\\
S_{\ell,\textrm{dS}}(n,\tau,N)&=S_{\ell,\textrm{dS},\infty}(n,\tau).
\end{align}
We point out again that we take into account angular-momentum sectors up to $\ell \sim 10^6$ in order to achieve the required accuracy of our calculation, even though the finite-size analysis is performed only for $\ell \leq 220$. To obtain results for the continuous theory in the infinite-size limit, we need the coefficients $S_{\ell,\textrm{flat},\infty}(n)$, $S_{\ell,\textrm{dS},\infty}(n,\tau)$ and $S_{\textrm{IR}}(n,\tau)$.
\item In order to calculate the sum of the contributions of all angular-momentum sectors, as in equation \eqref{eq:SEE_sum_dS}, we calculate the truncated sums
\begin{equation}\label{eq:trunc_sum_vac}
S_{\textrm{flat},\infty}(n;\ell_\textrm{max})=\sum_{\ell=0}^{\ell_\textrm{max}}(2\ell+1)S_{\ell,\textrm{flat},\infty}(n)
\end{equation}
and
\begin{equation}\label{eq:trunc_sum_dS}
S_{\textrm{dS},\infty}(n,\tau;\ell_\textrm{max})=\sum_{\ell=0}^{\ell_{\textrm{max}}}\left(2\ell+1\right)S_{\ell,\textrm{dS},\infty}(n,\tau)
\end{equation}
for various values of $\ell_\textrm{max}$. Then, we study their behaviour as functions of $\ell_\textrm{max}$ to obtain the limit $\ell_\textrm{max}\rightarrow\infty$, namely
\begin{equation}
S_{\textrm{flat},\infty}(n)=\lim_{\ell_\textrm{max}\rightarrow\infty}S_{\textrm{flat},\infty}(n;\ell_\textrm{max})
\end{equation}
and
\begin{equation}
S_{\textrm{dS},\infty}(n,\tau)=\lim_{\ell_\textrm{max}\rightarrow\infty}S_{\textrm{dS},\infty}(n,\tau;\ell_\textrm{max}).
\end{equation}
\item For the theory in flat space, according to \cite{Srednicki:1993im}, the asymptotic behaviour of $S_\ell(n)$ for $\ell\gg 1$ is given by
\begin{equation}
S_\ell(n)=\xi_\ell(n)[-\ln\xi_\ell(n)+1],
\end{equation}
where
\begin{equation}
\xi_\ell(n)=\frac{n(n+1)(2n+1)^2}{64\ell^2(\ell+1)^2}+\mathcal{O}(\ell^{-6}).
\end{equation}
This, in turn, implies that the truncated sum \eqref{eq:trunc_sum_vac}, behaves as
\begin{multline}
\label{eq:lmax_exp}
S_{\textrm{flat},\infty}(n;\ell_\textrm{max})=S_{\textrm{flat},\infty}(n)\\
+\sum_{i=1}^{i_\textrm{max}}\frac{1}{\ell_\textrm{max}^{2i}}(a_i(n)+b_i(n)\ln{\ell_{\textrm{max}}}).
\end{multline}
It turns out that for de Sitter space, for which \eqref{eq:trunc_sum_dS} applies, we have the same expansion, namely
\begin{multline}
\label{eq:lmax_exp_dS}
S_{\textrm{dS},\infty}(n,\tau;\ell_\textrm{max})=S_{\textrm{dS},\infty}(n,\tau)\\+\sum_{i=1}^{i_\textrm{max}}\frac{1}{\ell_\textrm{max}^{2i}}(a_i(n,\tau)+b_i(n,\tau)\ln{\ell_{\textrm{max}}}).
\end{multline}
\item For de Sitter space, our data span instants in time from $\tau=-3000$ to $\tau=-300$ in steps of $50$.  These are sufficiently early times so that the perturbative expansion of section \ref{subsec:expansion} is applicable. In accordance with the analysis there, we fit $S_{\textrm{IR}}(n,\tau)$ and $S_{\infty}(n,\tau)$ through a series in $1/\tau^2$, namely 
\begin{align}
S_{\textrm{IR}}(n,\tau)&=\sum_{i=0}^{i_{\textrm{max}}}\frac{S^{(2i)}_\textrm{IR}(n)}{\tau^{2i}}, \label{eq:tau_series_IR}\\
S_{\textrm{dS},\infty}(n,\tau)&=\sum_{i=0}^{i_{\textrm{max}}}\frac{S^{(2i)}_{\textrm{dS},\infty}(n)}{\tau^{2i}}.\label{eq:tau_series_inf}
\end{align}
As expected, we obtain $S^{(0)}_\textrm{IR}(n)=0$. 
\item Finally we analyze the dependence of $S_{\textrm{flat},\infty}(n)$, $S^{(i)}_\textrm{IR}(n)$ and $S^{(i)}_{\textrm{dS},\infty}(n)$ on $n$.
\end{enumerate}

\section{Summary of the results}
\label{sec:results}

In this section we summarize the results of our analysis. The details of the numerical calculations are presented in a series of appendices, to which we refer throughout this section. We consider the $(1+1)$- and the $(3+1)$-dimensional massless scalar theory  in flat and de Sitter space. The $(1+1)$-dimensional de Sitter theory corresponds to a toy model with a non-minimal coupling to gravity, described in \cite{Boutivas:2023ksg}. For the purposes of the current study, the details of the toy model are not important. The framework requires only the presence of a scalar field in $1+1$ dimensions, with its canonical modes at a state with  wave function given by \eqref{groundstatee}. In the context of the $(3+1)$-dimensional theory, this toy model describes the $\ell=0$ sector in an expansion in spherical harmonics.

For the $(1+1)$- and the $(3+1)$-dimensional massless scalar theory in flat space we calculated the coefficients of the universal logarithmic terms in \eqref{eq:SEE_expansion_flat} and  \eqref{eq:SEE_expansion_flat_1d}. We obtained
\begin{equation}\label{eq:c_d_values}
c = 0.999999999975, \quad -\frac{1}{d_1}= 89.999858,
\end{equation}
which are extremely close to the known values $1$ and $90$. The detailed analysis is presented in appendices \ref{sec:1d_flat} and \ref{sec:3d_flat}. The above values must coincide with the leading contributions to the entropy in de Sitter space in the limit $\tau\to -\infty$. For the $(1+1)$-dimensional toy model, by computing the term $S^{(0)}_{\textrm{dS},\infty}(n)$ in the expansion \eqref{eq:tau_series_inf} we obtained
\begin{equation}
c = 0.999999999699,
\end{equation}
which is in perfect agreement with \eqref{eq:c_d_values}. The details of the analysis are presented in appendix \ref{sec:1d_dS}. The $(1+1)$-dimensional toy model essentially describes the $\ell=0$ sector of the $(3+1)$-dimensional theory. In appendix \ref{sec:3d_dS} we present the calculation for the full theory by taking into account all the $\ell$-sectors. The term $S^{(0)}_{\textrm{dS},\infty}(n)$ takes the form \eqref{eq:SEE_expansion_flat} with 
\begin{equation}
-\frac{1}{d_1}=90.00089,
\end{equation}
in excellent agreement with \eqref{eq:c_d_values}.

Next we turn our attention to contributions that arise from the expansion of the background. The leading effect involves the length of the entire system and is given by the first term in \eqref{eq:dS_3d_N_corrections}. This term arises from the $\ell=0$ sector and appears both in the $(1+1)$- and the $(3+1)$-dimensional theory. Its calculation is presented in appendices \ref{sec:1d_dS} and \ref{sec:3d_dS}, respectively. Analyzing the terms $S_{\textrm{IR}}^{(2i)}(n)$ and neglecting contributions that are suppressed in the $(1+1)$-dimensional theory for $\epsilon \ll r$, or equivalently $n_r \gg 1$,  we obtain
\begin{equation}\label{eq:SEE_IR_2d}
S_{\textrm{IR}}^{(2)}(n)=c_{\textrm{IR}}^{(2)}n_r^2, \quad n_r=n+\frac{1}{2},
\end{equation}
where the value of the coefficient $c_{\textrm{IR}}^{(2)}$ is
\begin{equation}\label{eq:c_IR_2d}
c_{\textrm{IR}}^{(2)} = 0.3333366.
\end{equation}
We do not have sufficient precision to calculate reliably coefficients of higher-order terms in the $1/\tau^2$ expansion, but the plots on the second row of Figure \ref{fig:dS_2d_IR_terms} indicate that
\begin{equation}\label{eq:SEE_IR_n}
S_{\textrm{IR}}^{(2i)}(n)\sim n_r^{2i}.
\end{equation}
Recall that $\tau$ is measured in units of the UV cutoff. Thus, in the continuous theory $S_{\textrm{IR}}(r,\tau)$ admits an expansion of the form
\begin{equation}
	S_{\textrm{IR}}(r,\tau)=\sum_{i=1}^{\infty} c_{\textrm{IR}}^{(2i)}\left(\frac{r}{\tau}\right)^{2i},
\end{equation}
where $r=n_r\epsilon$ is the radius of the entangling surface. This is in line with the expansion \eqref{eq:SEE_expansion}. For the $(3+1)$-dimensional theory we obtain
\begin{equation}
S_{\textrm{IR}}^{(2)}(n)=c_{\textrm{IR}}^{(2)}n_R^2, \quad n_R=n+\frac{1}{2},
\end{equation}
where $c_{\textrm{IR}}^{(2)}$ is 
\begin{equation}\label{eq:c_IR_4d}
c_{\textrm{IR}}^{(2)}=0.3333364.
\end{equation}
Bear in mind that we use different discretization schemes in $1+1$ and $3+1$ dimensions. Thus, the agreement between \eqref{eq:c_IR_2d} and \eqref{eq:c_IR_4d} confirms that we obtain scheme-independent results for the continuous theory. We also verify that $S_{\textrm{IR}}(R,\tau)$ has a structure similar to \eqref{eq:SEE_IR_n}, see the second row of Figure \ref{fig:dS_3d_IR_terms}.

We next turn to the IR-independent part of entanglement entropy and study the term $S^{(2)}_{\textrm{dS},\infty}(n)$ in the  $(1+1)$-dimensional toy model. We find that in the continuous theory $(\epsilon\ll r)$
\begin{equation}\label{eq:SEE_2d}
S_{\textrm{dS},\infty}^{(2)}(r) = a^{(2)\prime}_{2} r^2 \ln \frac{r}{\epsilon} + a^{(2)}_2 r^2,
\end{equation}
where
\begin{equation}\label{eq:SEE_2d_fit}
a^{(2)\prime}_{2}= -0.3334008,\quad a^{(2)}_2 = -0.1679215.
\end{equation}
The value of $c_{\textrm{IR}}^{(2)}$, see equation \eqref{eq:c_IR_2d}, and that of $a^{(2)\prime}_{2}$, suggest that the entanglement entropy at order $1/\tau^2$ reads
\begin{equation}
\begin{split}
S^{(2)}_{\textrm{dS}}(r)&=S^{(2)}_{\textrm{IR}}(r)\ln\frac{L}{\epsilon}+S_{\textrm{dS},\infty}^{(2)}(r)\\
&=\frac{1}{3}r^2\ln\frac{L}{r}+ a^{(2)}_2 r^2.
\end{split}
\end{equation}
There is a cancellation of the contributions involving the UV cutoff. A possible explanation of this result is that the expanding background can only induce a UV divergence subleading to the  one already  present in the flat-space case. In $1+1$ dimensions, the leading divergence in flat space is logarithmic, so that the expansion of the background can only introduce UV finite corrections. However, this is not the case  in $3+1$ dimensions, where a new logarithmic UV divergence may appear.

In an accompanying publication \cite{Boutivas:2024sat} we obtain
\begin{equation}
S^{(2)}_{\textrm{dS}}(r)=r^{2}\left(\frac{1}{3}\ln\frac{L}{2\pi r}+\frac{4}{9}\right)
\end{equation}
analytically, thus confirming that the cancellation of the UV contributions is exact. We also show that $a^{(2)}_2$ is actually given by $a^{(2)}_2=-\frac{1}{3}\ln\left(2\pi\right)+\frac{4}{9}\simeq-0.16818$. The value obtained by the numerical calculation is very close to the exact one. Putting everything together, we conclude that the leading terms (apart from a constant) in the early-time expansion of the entanglement entropy are 
\begin{equation}
		S^{(1\textrm{d})}_{\textrm{dS}} = 
		\frac{1}{6}\ln \frac{r}{\epsilon}		
		+\frac{1}{3}\frac{r^2}{\tau^2}\ln\frac{L}{2\pi r}+\frac{4}{9}\frac{r^2}{\tau^2}+\dots
\label{1p1final}
\end{equation}
Appendix \ref{sec:1d_dS} contains the detailed analysis of the $(1+1)$-dimensional toy model.

Analyzing $S^{(2)}_{\textrm{dS},\infty}(n)$ in the $(3+1)$-dimensional theory we obtain
\begin{equation}
S_{\textrm{dS},\infty}^{(2)}(n) = a^{(2)\prime}_{2} n_R^2\ln n_R + a^{(2)}_2 n_R^2,\quad n_R=n+\frac{1}{2},
\end{equation}
where
\begin{equation}
 a^{(2)\prime}_{2}=0.000025,\quad a^{(2)}_2=-0.142650.
 \label{a2prim}
\end{equation}
The term $n_R^2\ln n_R$ is either vanishing or very suppressed. A purely numerical analysis cannot settle this issue. We also performed a fit on the $\ell=0$ sector and another fit on the sum of the rest of the $\ell$-sectors obtaining $-0.333498$ and $0.347750$ respectively. There is a discrepancy between $a^{(2)\prime}_{2}$ and the sum of these two numbers, which is equal to 0.014252. The numerical accuracy of our calculation is not sufficient for the precise determination of such a subleading term, especially if it has a very small coefficient. However, the presence of a new logarithmic UV-divergent term in the $(3+1)$-dimensional theory, in contrast to the  $(1+1)$-dimensional case, is a solid conclusion. The leading terms (apart from a constant) in the early-time expansion of the entanglement entropy are 
\begin{multline}
S^{(3\textrm{d})}_{\textrm{dS}} = d_2 \frac{R^2}{\epsilon^2} - \frac{1}{90} \ln \frac{R}{\epsilon} +\frac{1}{3}\frac{R^2}{\tau^2}\ln\frac{L}{\epsilon}\\+a^{(2)\prime}_{2} \frac{R^2}{\tau^2}\ln \frac{R}{\epsilon} + a^{(2)}_2 \frac{R^2}{\tau^2}+\dots\label{3p1final}
\end{multline}
with $d_2=0.295431454$ (see table \ref{Tab:vacuum_3d_fit}) and $a^{(2)\prime}_{2} $,
$a^{(2)}_2$ given by  \eqref{a2prim}.
The detailed analysis of the 
$(3+1)$-dimensional theory in de Sitter space is presented in appendix \ref{sec:3d_dS}.

\section{Conclusions}\label{conclusions}

Our final result \eqref{3p1final} provides an approximation to the general expression \eqref{eq:SEE_expansion} valid for subhorizon entangling surfaces with radii $R_p \ll 1/H$. We verified that the first term in \eqref{eq:SEE_expansion} vanishes in de Sitter space, i.e. $f_3(HR_p)=0$, so that no volume effect appears. We found that no new UV divergences $\sim \epsilon_p^{-2}$ and $\sim \epsilon_p^{-1}$ appear. This means that $f_1(HR_p)=0$, while $f_2(HR_p)$ retains its value in flat space, i.e. $f_2(HR_p)=d_2$. The effect of the expansion is apparent in the functions $f_0(HR_p)$ and $g(HR_p)$, which assume the form 
\begin{align}
	f_0(HR_p) &= d_0 + a^{(2)}_{2} H^2R_p^2, \\
	g(HR_p)&=-\frac{1}{90}+a^{(2)\prime}_{2} H^2R_p^2.
\end{align}
The most interesting part of our calculation concerns the novel term $h(HR_p)$, which was found to be
\begin{equation}
	h(HR_p)=\frac{1}{3}H^2R_p^2
\end{equation}
at this order. We note again the absence of terms involving $\ln(HR_p)$ in the various functions. The reason is that we expanded around the flat-space solution,
assuming $HR_p \ll 1$, which precludes the appearance of such terms. They are expected to be visible in the regime $HR_p =R/\tau=\mathcal{O}(1)$ or larger. 
For this reason, it is very interesting to study the late-time regime in the future in order to determine the complete structure of equation \eqref{eq:SEE_expansion}. However, several difficulties will have to be faced. In the late-time limit $|\tau|/\epsilon=(1/H)/(a\epsilon) =(1/H)/\epsilon_p\to 0$, the discretization of the theory is problematic, as the physical UV cutoff length becomes longer than the Hubble radius. Mathematically this limit is well defined, but the physical significance of its results is questionable. This means that one has to work at sufficiently high values of $|\tau|$ and increase $R$ by increasing the number of lattice points, so as to make $R/\tau$ large. This will make the numerical calculation much more demanding. Another problem is the absence of a concrete understanding of the expected form of the various functions appearing in \eqref{eq:SEE_expansion} at late times. Without some guidance, performing fits of the numerical data is very challenging. A perturbative expansion around $\tau=0$ is not very helpful
because the discretized theory is far removed from the continuous one in this regime, as we mentioned above.

Besides the calculation of the corrections to the flat-space entanglement entropy induced by the cosmological expansion, the most significant qualitative outcome of this work is that the entanglement entropy of scalar fields in de Sitter space depends on the size of the \emph{overall system}.\footnote{It is interesting to note that a similar conclusion can be reached in the context of the DS/dS correspondence \cite{Geng:2019bnn}: for matter satisfying the null energy condition, an observer in de Sitter space doesn't have enough information to reconstruct the whole spacetime and needs information from beyond the horizon.} Each field mode in the Bunch-Davies vacuum is described by a complex Gaussian wave function, which can be interpreted as a squeezed state for any instant in time. The emergence of the term that depends on the size of the overall system is a consequence of this fact. The wave function probes a larger part of the Hilbert space due to squeezing, which results in enhanced entanglement entropy. Moreover, the IR modes of the theory are the ones that get more squeezed by the expansion of the background. The eigenfrequencies of the modes of sectors with non-vanishing angular momentum are bounded from below by $\ell$, whereas the lowest eigenfrequencies of the sector with vanishing angular momentum are related to the size of the overall system. It is this effectively $(1+1)$-dimensional sector that generates the term with the dependence on the size of the overall system.

In the context of cosmology, the parameter $L$ would be fixed by the total size of a possibly finite universe. Under the more common assumption of a spatially flat infinite universe, a bound on $L$ can be derived by assuming that inflation had a finite duration. Let us assume that inflation started at some finite time $\tau_0$, with no prior squeezing of the wave function. Modes with eigenfrequencies larger than $1/\tau_0$ are subhorizon at the beginning of inflation, and accumulate all the squeezing induced by the ensuing expansion. Thus they are not severely modified relatively to the exact de Sitter scenario.  However, modes with eigenfrequencies smaller than $ 1/\tau_0$  are already superhorizon in the beginning of inflation, so that their total squeezing is smaller than that in the pure de Sitter scenario. It can be checked that these modes do not contribute significantly to the logarithmic term \cite{Boutivas:2024sat}. This means that $L$ can be identified roughly with the wavelength of the first mode that exited the horizon just after inflation started, which may extend far beyond the horizon today.

\begin{acknowledgments}
N. Tetradis would like to thank S. Floerchinger for very useful discussions.
\end{acknowledgments}

\appendix
\section{C++ code}
\label{sec:code}
In our calculation we have to scan for dozens of values of $N$ and $\tau$ in order to study finite-size effects. Then, we sum the contributions of hundreds of thousands of angular-momentum sectors for fixed $N_0=60$, yet for dozens of values of $\tau$. Moreover, as we perform many fits trying to track very subleading terms, it is necessary to have as high precision as possible. This in turn implies that we have to subtract finite-size corrections for many angular-momentum sectors. Given these facts, it is unfeasible to perform the calculation in a reasonable amount of time using commercial software, such as Wolfram Mathematica or MathWorks Matlab.

We implement our code in C++ using the package Eigen for linear algebra. Besides being much faster, since C++ is a low-level programming language, this setup allows us to use floats of 128-bit precision, which corresponds to 33--36 significant digits. Moreover, the code implements parallelization when performing calculations on different $\ell$-sectors. When studying finite-size corrections, which is done for each $\ell$-sector separately, we parallelize the loop iterating over time instants. For flat space, when running on a laptop with Intel i7-7700HQ CPU @ 2.8~GHz 4 Cores/8 threads and 12.0~GB of DDR3 RAM using 7 threads, it takes approximately 5 minutes to calculate the contributions of $10^4$ angular-momentum sectors. Memory management is also quite efficient. When studying the finite-size corrections of the $\ell=0$ sector using $N=5000$ for $n < 60$, only 400~MB are required. Such calculations with very large matrices take approximately 6 hours, when multiple independent computations run in parallel.

Regarding the precision of the calculation, we checked that for $N=60$, $\ell=0$ and various time instants, our results for $S_{\ell=0}$ and those obtain by Mathematica using working precision of more than 300 significant digits, agree within about $30$ significant digits for any $n$ and $\tau$. For $\ell=10^4$ the results match for up to $25$ significant digits. For $\ell = 10^6$ they match to around $10$ significant digits. Since the contribution of these large-angular-momentum sectors to the entanglement entropy  is several orders of magnitudes more suppressed than that of the $\ell=0$ sector, the reduction of precision at large $\ell$ does not affect our results. We  estimate that we have a precision of at least 20 significant digits after accounting for all angular-momentum sectors.

\section{Numerical analysis: $(1+1)$-dimensional theory in flat space}
\label{sec:1d_flat}
\begin{figure*}[t]
\begin{picture}(71,40)
\put(6,0){\includegraphics[angle=0,width=0.65\textwidth]{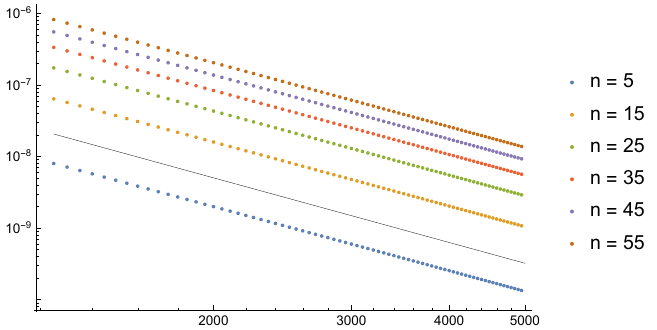}}
\put(59.5,1){$N$}
\put(0.5,30){$\frac{\Delta S_{\textrm{flat}}}{\Delta N}$}
\end{picture}
\caption{The finite difference $(S_{\textrm{flat}}(n,N+50)-S_{\textrm{flat}}(n,N))/{50}$, which approximates the $N$-derivative of $S_{\textrm{flat}}(n,N)$, for various values of $n$. For comparison, the solid black line has slope $-3$. It is evident that the dominant finite-size corrections to $S_{\textrm{flat},\infty}(n)$ scale as $N^{-2}$. A linear fit for each value of $n$ verifies that the slope is approximately $-2.97$ for all $n$. The small deviation at the third significant digit is due to subleading corrections.}\label{fig:vacuum_1d_1N_corrections}
\end{figure*}

In this appendix we describe the  computation of entanglement entropy in the $(1+1)$-dimensional massless field theory. First we subtract the finite-size effects.  Since the $(1+1)$-dimensional theory coincides with the $\ell=0$ sector of the $(3+1)$-dimensional one, we expect that the leading corrections scale as $N^{-2}$ \cite{Lohmayer:2009sq}. Figure \ref{fig:vacuum_1d_1N_corrections} verifies that the rate of change of $S_{\textrm{flat}}(n,N)$ with respect to $N$ is proportional to $N^{-3}$ for any $n$, to a good approximation. This implies that the dominant correction to $S_{\textrm{flat},\infty}(n)$ scales as $N^{-2}$. Of course, there are subleading corrections too. Determining them precisely results in a more accurate estimate of $S_{\textrm{flat},\infty}(n)$, and this is where our high-precision code comes in handy. Implementing the strategy described in section \ref{subsec:Our_Methodology} we conclude that the next-to-leading correction is proportional to $N^{-3}$. The full form of the finite-size corrections is given in \eqref{eq:vac_3d_N_corrections} for $\ell=0$. Dropping the subscript for the angular momentum, which is irrelevant for the $(1+1)$-dimensional theory, the expansion for the finite-size corrections assumes the form
\begin{equation}\label{eq:vac_1d_Nsq_corrections}
S_{\textrm{flat}}(n,N)=S_{\textrm{flat},\infty}(n)+\sum_{k=0}^{k_{\textrm{max}}}\frac{S_{\textrm{flat}}^{(k)}(n)}{N^{(k+2)}}.
\end{equation}
To carry out our calculation we perform a fit of the form \eqref{eq:vac_1d_Nsq_corrections} with $k_{\textrm{max}}=9$.  In total we fit 11 coefficients, but we are only interested in $S_{\textrm{flat},\infty}(n)$. When extrapolated to the limit $N\to \infty$, the data depend only on the dimensionless parameter $n$, which is related to the size of the subsystem as 
\begin{equation}
\label{eq:size_subsystem}
r =n_r\epsilon,\quad n_r= n+\frac{1}{2}.
\end{equation}
We then fit the data through an expansion whose leading term is of the form \eqref{eq:SEE_expansion_flat_1d}:
\begin{equation}\label{eq:vacuum_1d_SEE}
S_{\textrm{flat},\infty}\left(n\right)=\frac{c}{6}\ln n_r+d+ \sum_{i = 1}^{i_{\textrm{max}}} \frac{a_i}{n_r^{2i}} .
\end{equation}
The entanglement entropy of the continuous theory is determined only by the first two terms. It is well known that the central charge $c$ for the theory we are considering is equal to $1$, whereas the value of $d$ is scheme dependent. We verify this result with very high accuracy. Table \ref{Tab:vacuum_1d_fit} contains the values of the parameters $c$ and $d$ for a varying number of subleading corrections. By considering a large number of corrections, deviations from the exact value for $c$ are limited beyond the 10th significant digit. Of course, beyond some point we are restricted by the precision of the numerical calculation, the accuracy of the previous fits and the number of available data points. This is the reason why we quote in \eqref{eq:c_d_values} the value obtained when the dependence of $c$ on the number of correction terms reaches an extremum. This is the value closest to $1$, and is obtained when including 7 subleading terms. 
\begin{table}
\center
\begin{tabular}{|c | c | c|}
\hline
 $i_{\textrm{max}}$ & c & d \\
\hline
0 & 0.996167374949 & 0.0879500214016 \\
\hline
1 & 0.999869687409 & 0.0857936992834 \\
\hline
2 & 0.999995435360 & 0.0857176418509 \\
\hline
3 & 0.999999791602 & 0.0857149233720 \\
\hline
4 & 0.999999992458 & 0.0857147945304\\
\hline
5 & 0.999999999600 & 0.0857147898344\\
\hline
6 & 0.999999999968 & 0.0857147895867\\
\hline
7 & 0.999999999975 & 0.0857147895822\\
\hline
8 & 0.999999999921 & 0.0857147896201\\
\hline
\end{tabular}
\caption{The central charge $c$ and the constant term $d$ of the entanglement entropy \eqref{eq:SEE_expansion_flat_1d} of the $(1+1)$-dimensional massless theory when considering a variable number of subleading corrections in \eqref{eq:vacuum_1d_SEE}. We verify the known value $c=1$ with great accuracy. \label{Tab:vacuum_1d_fit}}
\end{table}

\section{Numerical analysis: $(3+1)$-dimensional theory in flat space}
\label{sec:3d_flat}
In this appendix we repeat the calculation of \cite{Lohmayer:2009sq} using the modifications described in section \ref{subsec:Our_Methodology} and the strategy discussed in section \ref{subsec:Summary_Numerical}. Since we are interested in performing next a much more involved calculation in a de Sitter background, we use the flat-space case in order to validate our code and the analysis strategy. In the process we re-derive known results with greater accuracy. Our goal is to establish a precision of at least $10^{-20}$ in the data after adding up all the angular-momentum sectors, as explained in appendix \ref{sec:code}.

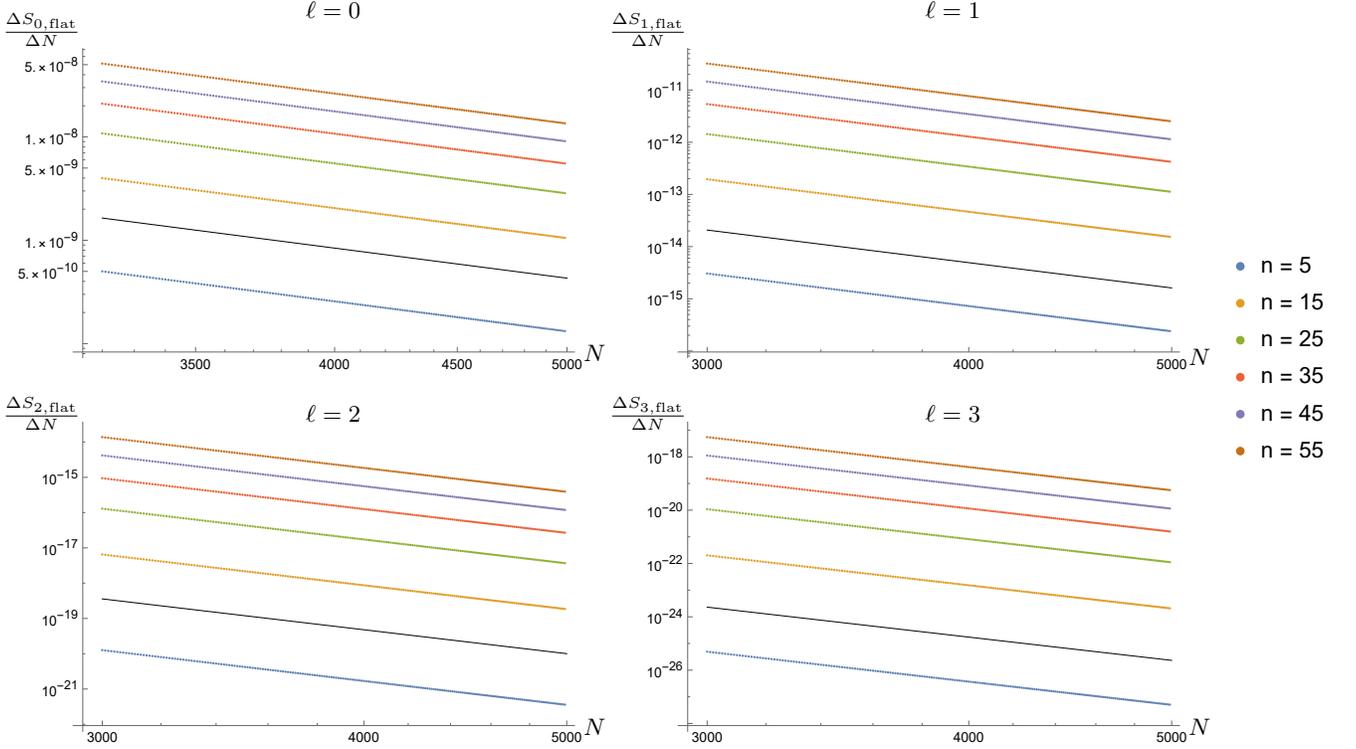
\begin{figure*}[t]
\centering
\begin{picture}(100,57)
\put(1,0){\includegraphics[angle=0,width=0.87\textwidth]{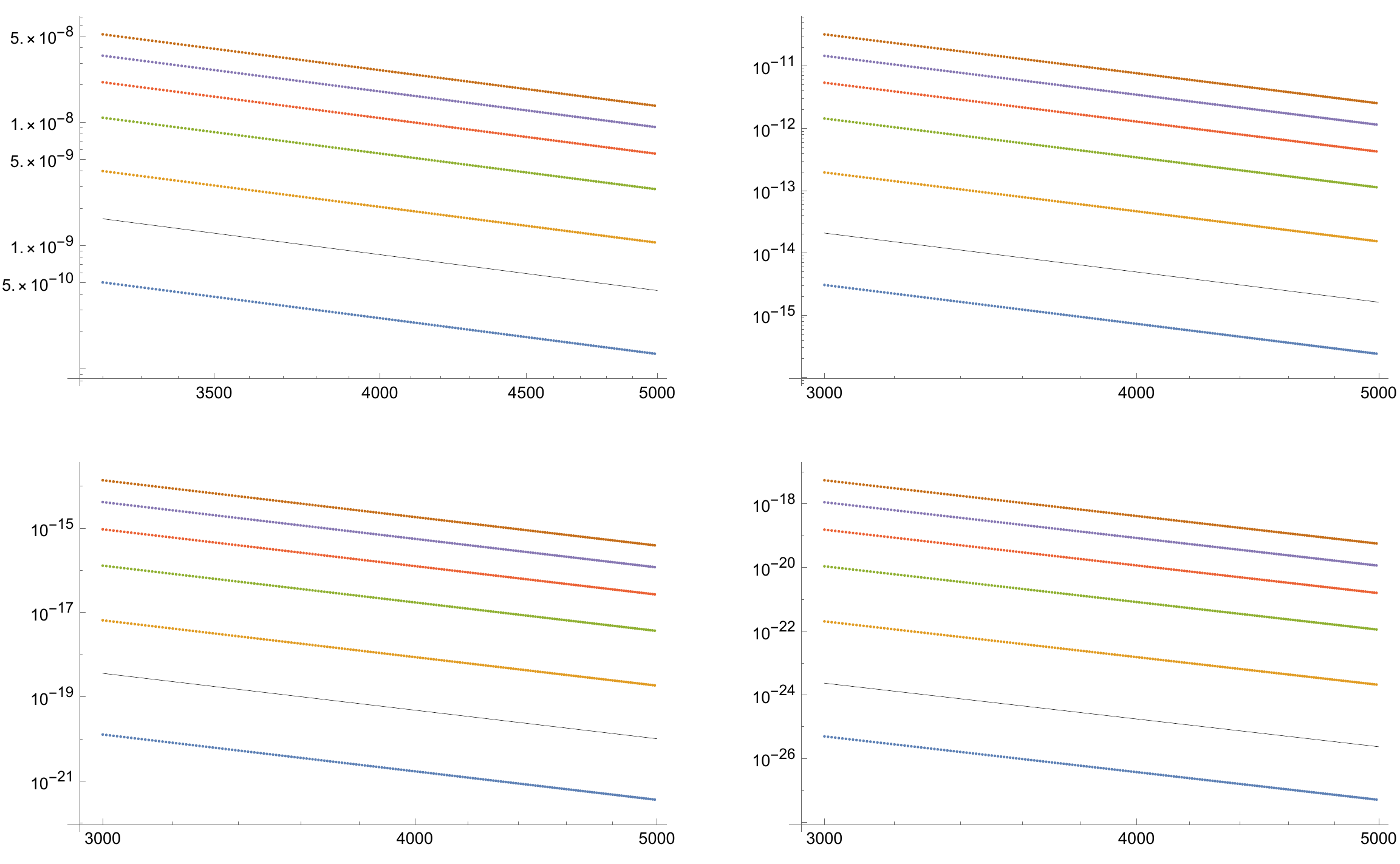}}
\put(22.5,25){$\ell=2$}
\put(22.5,55){$\ell=0$}
\put(68.5,25){$\ell=3$}
\put(68.5,55){$\ell=1$}
\put(88,1.75){$N$}
\put(88,29.25){$N$}
\put(43.12,1.75){$N$}
\put(43.12,29.5){$N$}
\put(0,53.5){$\frac{\Delta S_{0,\textrm{flat}}}{\Delta N}$}
\put(45,53.5){$\frac{\Delta S_{1,\textrm{flat}}}{\Delta N}$}
\put(0,25){$\frac{\Delta S_{2,\textrm{flat}}}{\Delta N}$}
\put(45,25){$\frac{\Delta S_{3,\textrm{flat}}}{\Delta N}$}
\put(90.5,21){\includegraphics[angle=0,width=0.085\textwidth]{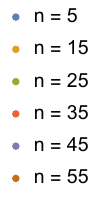}}
\end{picture}
\caption{The finite difference $({S_{\ell,\textrm{flat}}(n,N+10)-S_{\ell,\textrm{flat}}(n,N)})/{10}$, which approximates the $N$-derivative of $S_{\ell,\textrm{flat}}(n,N)$,  for $\ell=0$ to $\ell=3$ and various values of $n$. For comparison, the solid black lines in each plot have slope $-(2\ell+3)$.  It is evident that all lines are parallel, implying that the dominant finite size corrections scale as $N^{-(2\ell+2)}$. A linear fit for each value of $n$ verifies that the slopes are approximately $-2.996$, $-4.992$, $-6.989$, $-8.986$ for any $n$ and $\ell=0,1,2,3$, respectively. The small deviations are due to subleading corrections.}
\label{fig:vacuum_3d_1N_corrections}
\end{figure*}

We first have to subtract the finite-size effects. The entanglement entropy of any $\ell$-sector is given by \eqref{eq:vac_3d_N_corrections}. This allows us to the estimate $S_{\ell,\textrm{flat},\infty}(n)$ with high accuracy through a numerical fit, as long as we compute the entropy of each angular-momentum sector for large enough $N$. The fact that the dominant correction to $S_{\ell,\textrm{flat},\infty}(n)$ scales like $N^{-2(\ell+1)}$ can be verified through Figure \ref{fig:vacuum_3d_1N_corrections}. It can be seen that, for any $n$, the rate of change of $S_{\ell,\textrm{flat}}(n,N)$ with $N$ is proportional to $N^{-(2\ell+3)}$ from $\ell=0$ to $\ell=3$. We have confirmed that this behaviour persists all the way up to $\ell=220$, which is the last $\ell$-sector for which we take finite-size corrections into account. As previously discussed, the corrections for higher angular momenta are insignificant with our desired accuracy. For each $\ell$, apart from the leading term in $N$, we must also account for enough subleading corrections in order to obtain $S_{\ell,\infty}(n)$ with the highest possible accuracy. The range and the accuracy of our data allow for numerical fits with $10$ subleading corrections for the most important $\ell=0$ to $\ell=4$ sectors, and with $3$ subleading corrections for the remaining ones ($\ell=5$ to $\ell=220$).

\begin{figure*}[t]
\centering
\begin{picture}(60,37)(0,0)
\put(0,0){\includegraphics[width=0.55\textwidth]{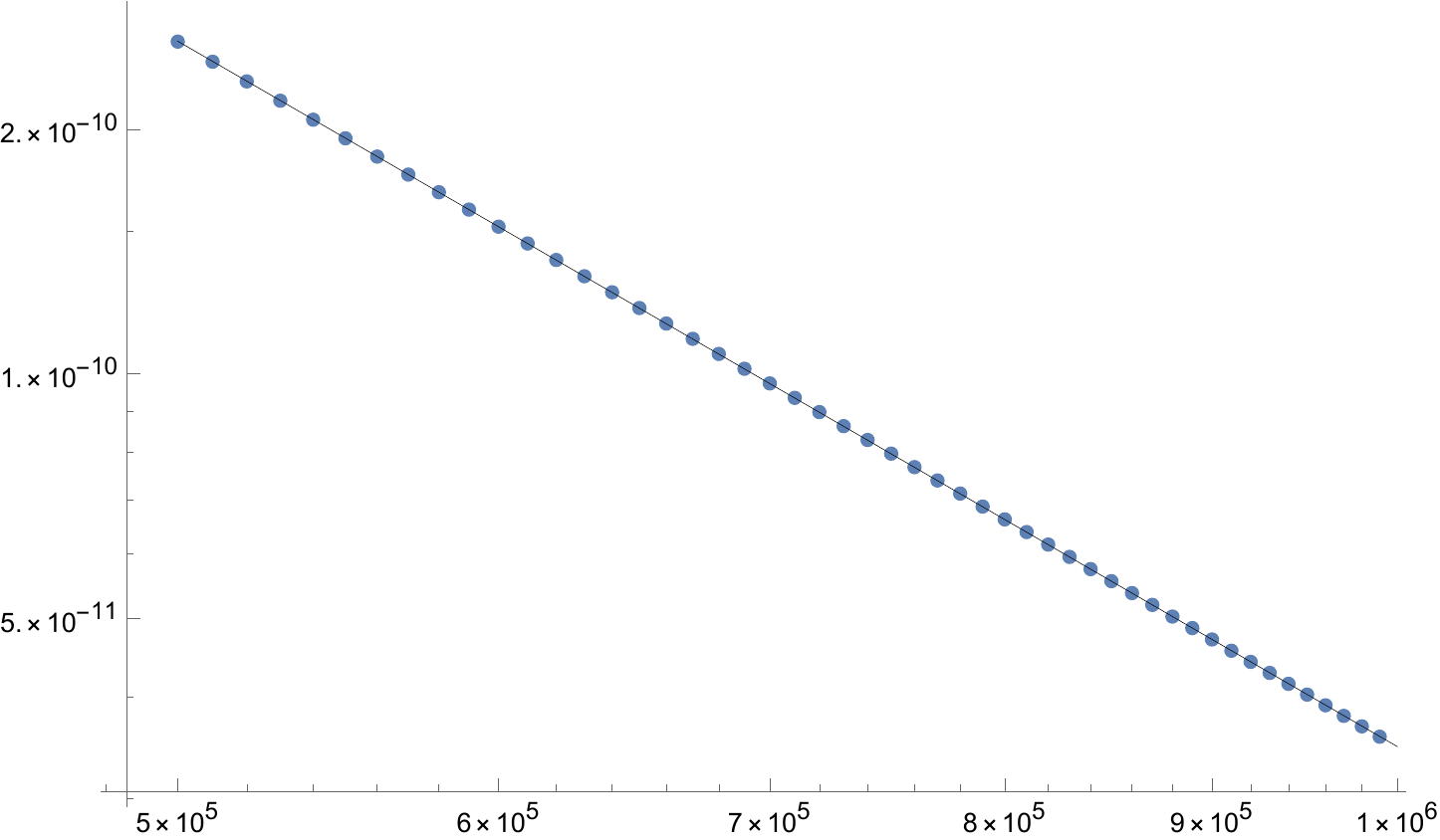}}
\put(55.5,1){$\ell_{\textrm{max}}$}
\put(1,33.5){$\frac{\Delta S_{\textrm{flat}}}{\Delta\ell_{\textrm{max}}}$}
\end{picture}
\caption{The finite difference $({S_{\textrm{flat}}(n,\ell_\textrm{max}+10^4)-S_{\textrm{flat}}(n,\ell_\textrm{max})})/{10^4}$ for $n=50$ and various values of $\ell_\textrm{max}$. A linear fit verifies that the slope is approximately $-2.885$. The small deviation is due to subleading corrections.}
\label{fig:lmax_limit_vac}
\end{figure*}

The next step towards the continuous theory involves the calculation of the $\ell_\textrm{max}\rightarrow\infty$ limit. Through a straightforward approach, we study the behaviour of the truncated sum $S_{\textrm{flat},\infty}(n;\ell_\textrm{max})$, see \eqref{eq:trunc_sum_vac}, and estimate its asymptotic value for $\ell_\textrm{max}\rightarrow\infty$ using a fit. The fact that the dominant correction of the truncated sum scales as $\ell_\textrm{max}^{-2}$ can be verified through Figure \ref{fig:lmax_limit_vac}. It can be seen that the rate of change of $S_{\textrm{flat},\infty}(n;\ell_\textrm{max})$ with $\ell_\textrm{max}$ is proportional to $\ell_\textrm{max}^{-3}$. However, as \eqref{eq:lmax_exp} suggests, subleading terms, both polynomial and logarithmic, must also be taken into account in the numerical fit in order to extract the asymptotic value with greater accuracy. Our data allowed for a fit with $10$ polynomial and $10$ logarithmic corrections. As for how the fit itself is performed, there are two alternative approaches: one could examine the behaviour of the truncated sum by progressively adding angular momenta as large sets, or simply one by one. In theory, the second option seems preferable. Nonetheless, one must consider that for large $\ell$'s the numerical accuracy drops, however insignificant their contribution to the overall sum may be. Therefore, in practice, we have verified that the first option leads to more accurate results regarding the asymptotic value of the sum for $\ell_\textrm{max}\rightarrow\infty$.

We have finally reached the point where our data depend only on the dimensionless parameter $n$, which characterizes the size of the subsystem, according to \eqref{eq:size_subsystem}. For the $(3+1)$-dimensional theory we use the notation $n_R=n+1/2$ instead of $n_r$  that we use in the $(1+1)$-dimensional case in order to help the reader to distinguish between the two cases. We are now able to fit the data to an expansion of the form \eqref{eq:SEE_expansion_flat} with the explicit addition of the subleading corrections:
\begin{equation}
\label{eq:vac_exp}
S_{\textrm{flat},\infty}(n) = d_2 n_R^2 + d_1 \ln n_R + d_0+\sum_{i=1}^{i_\textrm{max}}\frac{a_i}{n_R^{2i}}.
\end{equation}
Only the three first terms contribute to the entanglement entropy of the continuous theory. The first term is the famous area law, whose coefficient is scheme dependent, as is the case for the constant $d_0$. According to \cite{Solodukhin:2008dh,Casini:2009sr}, and as verified by \cite{Lohmayer:2009sq}, we expect $d_1=-{1}/{90}$. Our code allows the verification of this result with high accuracy. Table \ref{Tab:vacuum_3d_fit} contains the values of the parameters $d_2$,  $d_0$ and $-1/d_1$,  for various values of $i_\textrm{max}$. It is shown that more and more corrections improve the result, until stability is achieved. However, the inclusion of additional subleading terms does not lead to an improvement of the results beyond  
a certain point. This is due to the limited, albeit high, precision of the numerical calculation, the accuracy of the previous fits and the number of the available data points. In \eqref{eq:c_d_values} we quote the value closest to $-1/90$, which is obtained when including 6 subleading terms.

\begin{table}
\center
\begin{tabular}{|c | c | c| c |}
\hline
$i_\textrm{max}$ & $d_2$ & $d_0$ & $-1/d_1$\\
\hline
0 & 0.295431434011 & -0.035537375 & 90.60163\\
\hline
1 & 0.295431448040 & -0.035395550 & 90.24163\\
\hline
2 & 0.295431452662 & -0.035325992 & 90.07810 \\
\hline
3 & 0.295431453497 & -0.035309276 & 90.04061 \\
\hline
4 & 0.295431453836 & -0.035300798 & 90.02222\\
\hline
5 & 0.295431454029 & -0.035294952 & 90.00988\\
\hline
6 & 0.295431454164 & -0.035290102 & 89.99986\\
\hline
7 & 0.295431454273 & -0.035285571 & 89.99068\\
\hline
8 & 0.295431454369 & -0.035281006 & 89.98158\\
\hline
\end{tabular}
\caption{The area law coefficient $d_2$, the constant term $d_0$ and the denominator of the subleading logarithmic term of $S^{(3\textrm{d})}_{\textrm{flat}}=d_2n_R^2+d_1\ln n_R+d_0$ for a varying number of subleading corrections. We verify the known value $d_1=-1/90$ with high accuracy. \label{Tab:vacuum_3d_fit}}
\end{table}

\section{Numerical analysis: $(1+1)$-dimensional theory in de Sitter space}
\label{sec:1d_dS}
In this appendix we study the $(1+1)$-dimensional scalar field theory in de Sitter space. In $1+1$ dimensions, if one assumes a background given by the FRW metric
\begin{equation}
ds^2 = a^2 \left( \tau \right) \left( d\tau^2 - dr^2\right),
\end{equation}
the scalar field is canonically normalized. As a result, when considering a minimally coupled scalar field, the time-dependent mass term that emerges in $3+1$ dimensions is absent. As has been discussed in \cite{Boutivas:2023ksg}, the background has a non-trivial effect on the ground state of the theory only in the presence of a non-minimal coupling of the field to gravity. The inclusion of an effective mass term that depends on the background can be achieved by allowing for a non-minimal coupling to gravity of the form $\xi R \phi^2$. For a dS background with $a(\tau)=-1/(H\tau)$, the curvature scalar $R$ in $1+1$ dimensions is equal to $-2H^2$. The choice $\xi=-1/2$ results in an effective mass term $-2/\tau^2$ for the field. This leads to the construction of the toy model analyzed in \cite{Boutivas:2023ksg}, for which the ground state has the  same wave function \eqref{groundstatee} as for a minimally coupled field in $3+1$ dimensions. In the context of the $(3+1)$-dimensional theory that we consider in the following appendix, the present toy model describes the $\ell=0$ sector in an expansion in spherical harmonics. In this sense, the results of this appendix are also relevant for the $(3+1)$-dimensional case, in which the contributions from all values of $\ell$ must be added up.

\begin{figure*}[t]
\centering
\begin{picture}(96,60)
\put(2,29){\includegraphics[angle=0,width=0.42\textwidth]{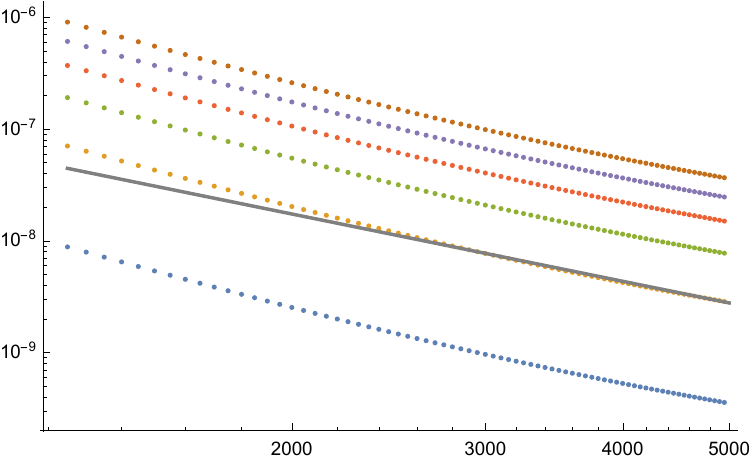}}
\put(20,56){$\tau=-3000$}
\put(44,30){$N$}
\put(1,56.25){$\frac{\Delta S_{\textrm{dS}}}{\Delta N}$}
\put(50,29){\includegraphics[angle=0,width=0.44\textwidth]{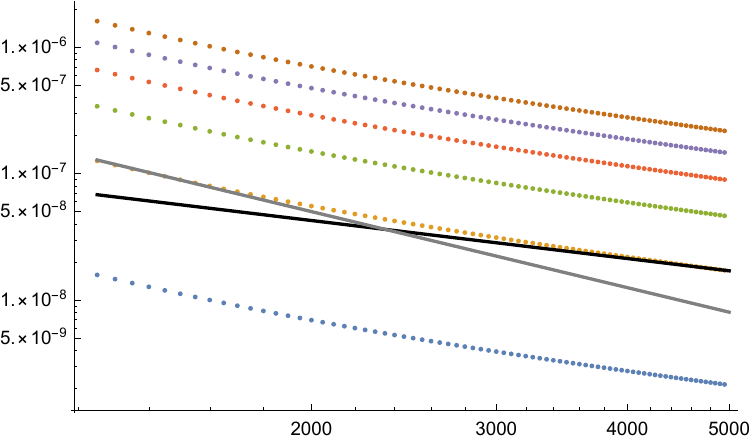}}
\put(70,56){$\tau=-1000$}
\put(94,30){$N$}
\put(51,56.25){$\frac{\Delta S_{\textrm{dS}}}{\Delta N}$}
\put(0,0){\includegraphics[angle=0,width=0.44\textwidth]{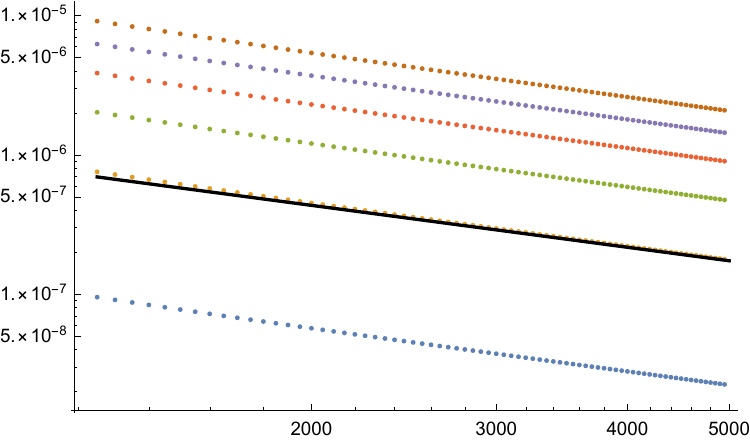}}
\put(21,26){$\tau=-300$}
\put(44,1){$N$}
\put(1,27.25){$\frac{\Delta S_{\textrm{dS}}}{\Delta N}$}
\put(55,6){\includegraphics[angle=0,width=0.085\textwidth]{dS_legendV}}
\end{picture}
\caption{The finite difference $({S_{\textrm{dS}}(n,\tau,N+50)-S_{\textrm{dS}}(n,\tau,N)})/{50}$ which approximates the $N$-derivative of $S_{\textrm{dS}}(n,\tau,N)$, for various values of $n$ and $\tau$. For comparison, the solid black lines  have slope $-1$, while the gray ones  have slope $-2$. For fixed $\tau$ all lines exhibit similar behaviour. The form of the dominant finite-size effect depends on $\tau$. At $\tau \to -\infty$  the leading finite-size correction to the entropy scales as $N^{-2}$, as deduced from Figure \ref{fig:vacuum_1d_1N_corrections}. For growing $\tau$, the dominant correction has a softer dependence on $N$, and eventually becomes proportional to $\ln N$.}
	\label{fig:dS_1d_1N_corrections}
\end{figure*}

We first have to isolate the terms of entanglement entropy that survive when the size of the system is infinite. The entanglement entropy is a function of $3$ parameters, and we denote it as $S_{\textrm{dS}}(n,\tau,N)$. The main difficulty in analyzing the data is that we do not have a priori information on the form of the finite-size corrections to guide us through the numerical analysis. The problem is more pronounced at late times, when the system deviates more strongly from the behaviour typical of a flat background. In order to overcome this difficulty we limit our analysis to sufficiently early times ($|\tau| \gg 1$), during which the perturbative expansion in $1/\tau^2$ developed in section  \ref{subsec:expansion} can provide useful information. We perform our analysis for $\tau \leq -300$. This choice ensures that de Sitter effects are suppressed enough so that  only a few terms in the $1/\tau^{2}$ expansion contribute significantly to the entanglement entropy, yet their effect is still visible.

We start by studying the behaviour of $S_{\textrm{dS}}(n,\tau,N)$ as a function of $N$ for fixed $n$ and $\tau$. We analyze the rate of change of $S_{\textrm{dS}}(n,\tau,N)$ with respect to $N$ for some indicative times. In Figure \ref{fig:dS_1d_1N_corrections} we present this rate of change for $\tau=-3000$, $\tau=-1000$ and $\tau=-300$. We already know that for $\tau=-\infty$ the slope is approximately $-3$, as in this limit the entropy approaches its form in a flat background. At later times there is a transition period, with softer slopes that vary in different ranges of $N$. Finally, the slope becomes approximately $-1$ for $\tau \simeq -300$. It is clear that different forms of corrections to the entropy are dominant at different times $\tau$. Initially the dominant correction is $N^{-2}$, as deduced from Figure \ref{fig:vacuum_1d_1N_corrections} for the flat case. For $\tau = -3000$ the dependence on $N$ becomes softer. For $\tau = -1000$  we see that for small values of $N$ the corrections to the entropy scale as $N^{-1}$, while for large values of $N$ the term $\ln N$ is dominant. Finally, for $\tau = -300$ the term $\ln N$ is dominant for any value of $N$ within the range that we consider. 

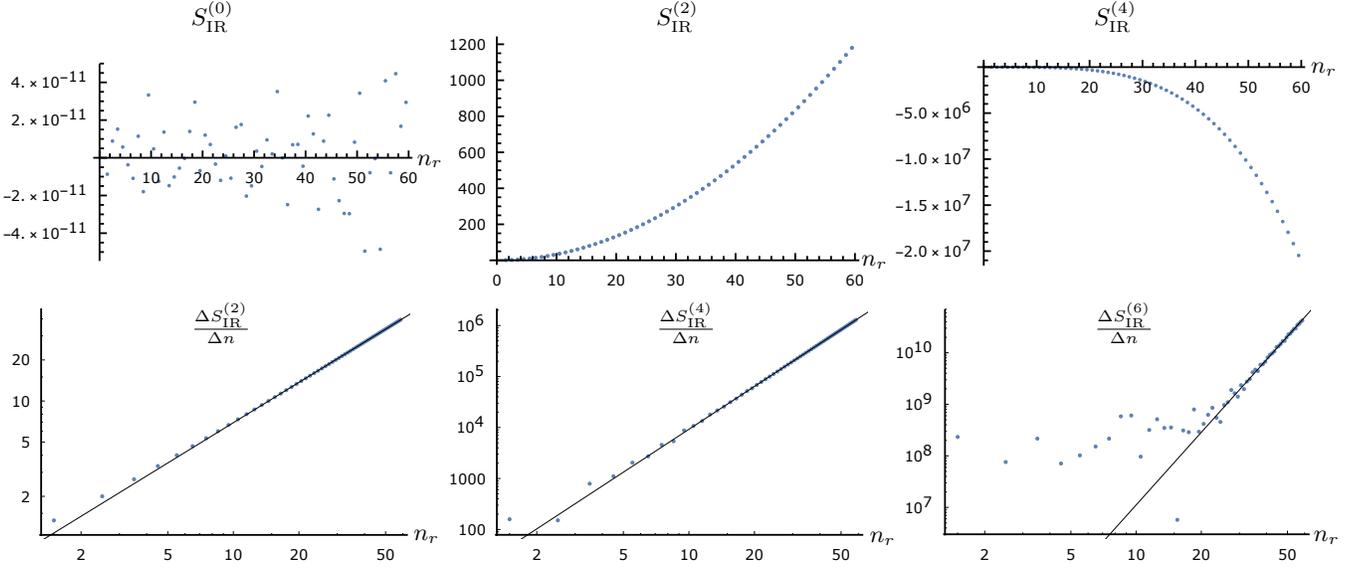
\begin{figure*}[t]
\centering
\begin{picture}(100,43)
\put(0,0){\includegraphics[angle=0,width=0.97\textwidth]{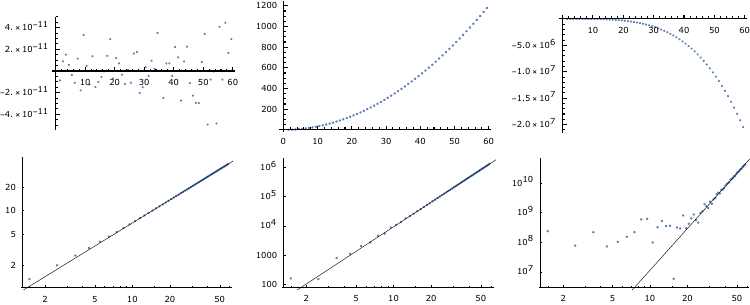}}
\put(14,40){$S_{\textrm{IR}}^{(0)}$}
\put(48.5,40){$S_{\textrm{IR}}^{(2)}$}
\put(81,40){$S_{\textrm{IR}}^{(4)}$}
\put(14,17){$\frac{\Delta S_{\textrm{IR}}^{(2)}}{\Delta n}$}
\put(48.5,17){$\frac{\Delta S_{\textrm{IR}}^{(4)}}{\Delta n}$}
\put(81,17){$\frac{\Delta S_{\textrm{IR}}^{(6)}}{\Delta n}$}
\put(30.5,1.75){$n_r$}
\put(64.35,1.6){$n_r$}
\put(97.3,1.7){$n_r$}
\put(30.75,29.65){$n_r$}
\put(63.75,22.15){$n_r$}
\put(97.1,36.5){$n_r$}
\end{picture}
\caption{  The first row depicts $S_{\textrm{IR}}^{(0)}(n)$ (left panel), $S_{\textrm{IR}}^{(2)}(n)$ (middle panel) and $S_{\textrm{IR}}^{(4)}(n)$ (right panel). $S_{\textrm{IR}}^{(0)}(n)$ appears as random noise of very small amplitude, in accordance with the fact that no such term is expected in the flat space limit. The second row depicts the finite differences $\vert S_{\textrm{IR}}^{(2)}(n+1) -S_{\textrm{IR}}^{(2)}(n) \vert$ (left panel), $\vert S_{\textrm{IR}}^{(4)}(n+1) -S_{\textrm{IR}}^{(4)}(n) \vert$ (middle panel) and $\vert S_{\textrm{IR}}^{(6)}(n+1) - S_{\textrm{IR}}^{(6)}(n) \vert$ (right panel). The solid black lines are linear fits having slope $0.98$, $2.79$ and $4.68$, respectively, implying that the leading behaviour of the corrections is of the form $\left(n_r/\tau\right)^{2i}$. Recall that $n_r=n+1/2$ and $r=n_r\epsilon$ in the continuous theory.}
\label{fig:dS_2d_IR_terms}
\end{figure*}

For large $N$  the entanglement entropy has an expansion of the form \eqref{eq:dS_3d_N_corrections} with $\ell=0$. Dropping the subscript for the angular momentum, which is irrelevant for the $(1+1)$-dimensional theory, the expansion for the finite-size corrections is given by 
\begin{multline}\label{eq:Large_N_fit}
S_{\textrm{dS}}(n,\tau,N)=S_{\textrm{IR}}(n,\tau)\ln N\\  +S_{\textrm{dS},\infty}(n,\tau)+\sum_{k=1}^{k_{\textrm{max}}}\frac{S_{\textrm{dS}}^{(k)}(n,\tau)}{N^{k}}.
\end{multline}
We perform a fit keeping corrections up to $k_{\textrm{max}}=9$. We are interested in the terms $S_{\textrm{IR}}(n,\tau)$ and $S_{\textrm{dS},\infty}(n,\tau)$, because they are the only ones that survive in the limit $L\gg R$. It must be emphasized that the first term in \eqref{eq:Large_N_fit} is always subleading to the second one for the parameter range that we consider in our analysis ($N\leq 5000$). In principle, one may consider larger values of $N$ in order to try to invert this relation. However, the  perturbative expansion of section \ref{subsec:expansion} fails in this case. 

We proceed by fitting $S_{\textrm{IR}}(n,\tau)$ and $S_{\textrm{dS},\infty}(n,\tau)$  using the series in $1/\tau^2$, see equations \eqref{eq:tau_series_IR} and \eqref{eq:tau_series_inf}. To validate our calculation we compare $S_{\textrm{dS},\infty}^{(0)}(n)$ with the flat-space results, presented in appendix \ref{sec:1d_flat}. Implementing a third fit of the form \eqref{eq:vacuum_1d_SEE}, including corrections up to $n_r^{-10}$, we obtain
\begin{equation}
c = 0.999999999699, \quad d = 0.0857147897934.
\end{equation}
These values are extremely close to the ones we obtained when analyzing the data in flat space, see table \ref{Tab:vacuum_1d_fit}.

Turning to $S_{\textrm{IR}}(n,\tau)$, the first row of Figure \ref{fig:dS_2d_IR_terms} depicts $S_{\textrm{IR}}^{(0)}(n)$, $S_{\textrm{IR}}^{(2)}(n)$ and $S_{\textrm{IR}}^{(4)}(n)$. It is clear that $S_{\textrm{IR}}^{(0)}(n)$ is vanishing, as expected, and the leading term is $S_{\textrm{IR}}^{(2)}(n)$. In order to identify the scaling of the corrections, in the second row of Figure \ref{fig:dS_2d_IR_terms} we plot the finite differences $\vert S^{(2i)}_\textrm{IR}(n+1) -S^{(2i)}_\textrm{IR}(n)\vert$ for $i=1,2,3$. For the term scaling as $S^{(6)}_\textrm{IR}(n)$ the first data points are not reliable enough to be included in the
 analysis. The solid black lines are linear fits having slopes $0.98$, $2.79$ and $4.68$, implying that the leading behaviour of the corrections is of the form $\left(n_r/\tau\right)^{2i}$, where $n_r=n+1/2$. Since we have reached the limit of the precision of our numerical calculation, and there are no analytical results available to compare with, we feel confident enough to provide a numerical estimate only for the coefficient of the term $\left(n_r/\tau\right)^{2}$. A fit of the form
\begin{equation}\label{eq:IR_2d_fit}
S_{\textrm{IR}}^{(2)}(n) = c_{\textrm{IR}}^{(2)} n_r^2 +c \ln n_r +d
\end{equation}
returns
\begin{equation}\label{eq:IR_2d_coef}
c_{\textrm{IR}}^{(2)} = 0.3333366.
\end{equation}
This value strongly indicates that the  continuum-limit, early-time expansion of the part of the entanglement entropy that depends on the size $L$ of the entire system reads
\begin{equation}\label{eq:2d_IR_div}
S_{\textrm{IR}}(r, \tau)=\frac{1}{3}\frac{r^2}{\tau^2}\ln\frac{L}{\epsilon}+\dots ,
\end{equation}
where we have reinstated a factor of the UV cutoff for dimensional reasons. It turns out that this expression can be confirmed through an analytical calculation that will be presented separately.

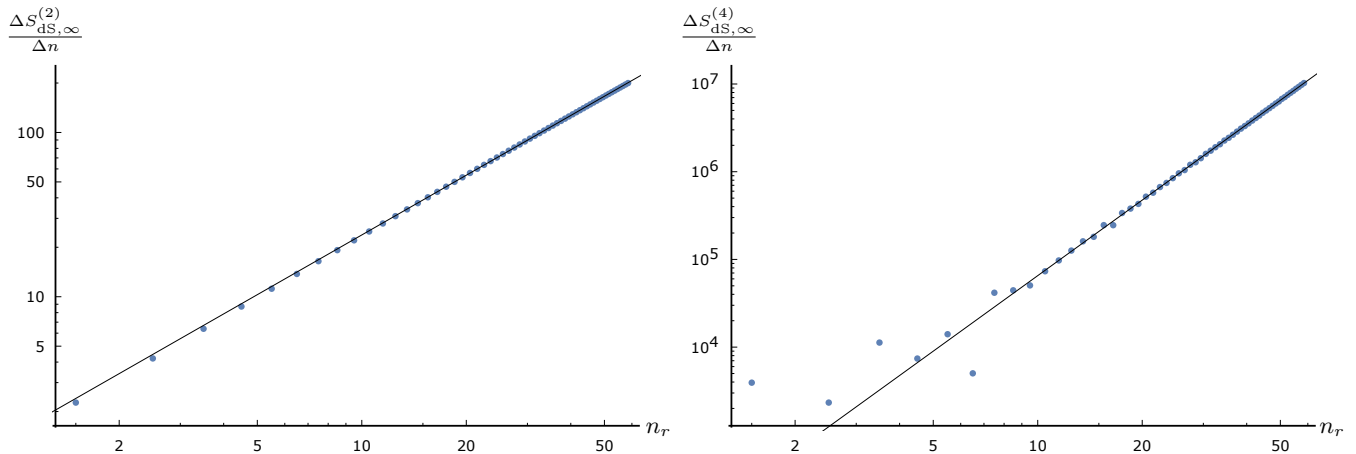
\begin{figure*}[t]
\centering
\begin{picture}(100,36)
\put(0.5,0){\includegraphics[angle=0,width=0.97\textwidth]{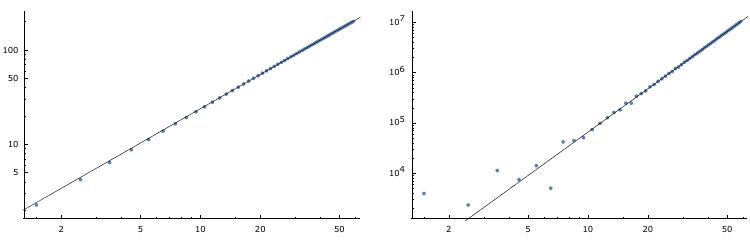}}
\put(47.5,3){$n_r$}
\put(97.5,3){$n_r$}
\put(0,32){$\frac{\Delta S_{\textrm{dS},\infty}^{(2)}}{\Delta n}$}
\put(50,32){$\frac{\Delta S_{\textrm{dS},\infty}^{(4)}}{\Delta n}$}
\end{picture}
\caption{The finite differences $\vert S_{\textrm{dS},\infty}^{(2)}(n+1) -S_{\textrm{dS},\infty}^{(2)}(n) \vert$ (left panel) and $\vert S_{\textrm{dS},\infty}^{(4)}(n+1) -S_{\textrm{dS},\infty}^{(4)}(n) \vert$ (right panel). The solid black lines have slope $1.21$ and $2.86$, respectively. It follows that $S_{\textrm{dS},\infty}^{(2)}(n)$ contains both $n_r^2 \ln n_r$ and $n_r^2$ terms. The precision of the data is insufficient for an accurate analysis of $S_{\textrm{dS},\infty}^{(4)}(n)$. Nevertheless, it is natural to assume that both $n_r^4\ln n_r$ and $n_r^4$ terms are present. Recall that $n_r=n+1/2$ and $r=n_r\epsilon$ in the continuous theory.}
\label{fig:dS_2d_finite_terms}
\end{figure*}

We next turn to $S_{\textrm{dS},\infty}(n,\tau)$ and analyze $S_{\textrm{dS},\infty}^{(2)}(n)$ and $S_{\textrm{dS},\infty}^{(4)}(n)$. Figure \ref{fig:dS_2d_finite_terms} contains the rate of change of these terms. In the case of $S_{\textrm{dS},\infty}^{(2)}(n)$ the slope is $1.21$, while for $S_{\textrm{dS},\infty}^{(4)}(n)$ the slope is $2.86$. Regarding $S_{\textrm{dS},\infty}^{(4)}(n)$, it is evident that we do not have enough precision for small $n$, and we will not analyze it any further. Since the slope of the change of  $S_{\textrm{dS},\infty}^{(2)}(n)$ is larger than $1$, there is not only a $n_r^2$ contribution, but a term $n_r^2\ln n_r$ as well. Performing a fit of the form
\begin{multline}\label{eq:finite_2d_fit}
S_{\textrm{dS},\infty}^{(2)}(n) = a^{(2)\prime}_{2} n_r^2\ln n_r + a^{(2)}_2 n_r^2 + c \ln n_r  \\+ d + \frac{a_{-2}}{n_r^2}+ \frac{a_{-4}}{n_r^4}+ \frac{a_{-6}}{n_r^6},
\end{multline}
we obtain
\begin{equation}
\label{eq:finite_2d_coefs}
\begin{split}
a^{(2)\prime}_{2}&= -0.3334008,\\
a^{(2)}_2 &= -0.1679215.
\end{split}
\end{equation}
 In \cite{Boutivas:2024sat} we show analytically that the exact values are 
 \begin{equation}\label{eq:finite_2d_coefs_exact}
 \begin{split}
 	a^{(2)\prime}_{2}&=-c_{\textrm{IR}}^{(2)}=-\frac{1}{3},\\
 	a^{(2)}_2 &= -\frac{1}{3}\ln\left(2\pi\right)+\frac{4}{9}\simeq-0.16818.
 	\end{split}
 \end{equation}
Putting everything together, for the continuous theory in the early-time expansion we obtain
\begin{multline}
S_{\textrm{IR}}(r, \tau)\ln\frac{L}{\epsilon}+S_{\textrm{dS},\infty}(r, \tau)\\= S_{\textrm{flat}}(r)+\frac{r^2}{\tau^2}\left(\frac{1}{3}\ln\frac{L}{2\pi r}+\frac{4}{9}\right)+\dots.
\end{multline}
It is remarkable that the various terms conspire so as to eliminate a dependence on the UV cutoff from the final expression  for the correction arising from the dS background. It turns out that this occurs only in $1+1$ dimensions, while a UV-dependent correction survives in $3+1$ dimensions. One can speculate that the expanding background can only induce a UV divergence subleading to the  one already  present in the flat-space case. In $1+1$ dimensions, the leading divergence in flat space is logarithmic, so that the expansion of the background can only introduce UV finite corrections. On the other hand, in $3+1$ dimensions, new UV divergences may appear, as is shown in the following appendix.

\section{Numerical analysis: $(3+1)$-dimensional theory in de Sitter space}
\label{sec:3d_dS}
We finally consider the $(3+1)$-dimensional scalar field theory in de Sitter space. The entanglement entropy is a function of three dimensionless parameters, and consequently we denote it as $S_\textrm{dS}(n,\tau,N)$, with $\tau$ expressed in units of the UV cutoff. As before, we work in a certain time regime, namely from $\tau=-3000$ to $\tau=-300$ in steps of $50$, during which the de Sitter effects are sufficiently suppressed compared to the flat-space effects for an expansion in $1/\tau^2$ to converge, but not too suppressed to be untraceable. As we discussed in appendix \ref{sec:1d_dS} for the $(1+1)$-dimensional toy model, the large-size limit is non-trivial because of the appearance of a term proportional to $\ln N$. Thus, we first study the entanglement entropy as a function of $N$ for fixed $n$ and $\tau$. 
 
\begin{figure*}[p]
\centering
\begin{picture}(100,57.5)
\put(0,0){\includegraphics[angle=0,width=0.88\textwidth]{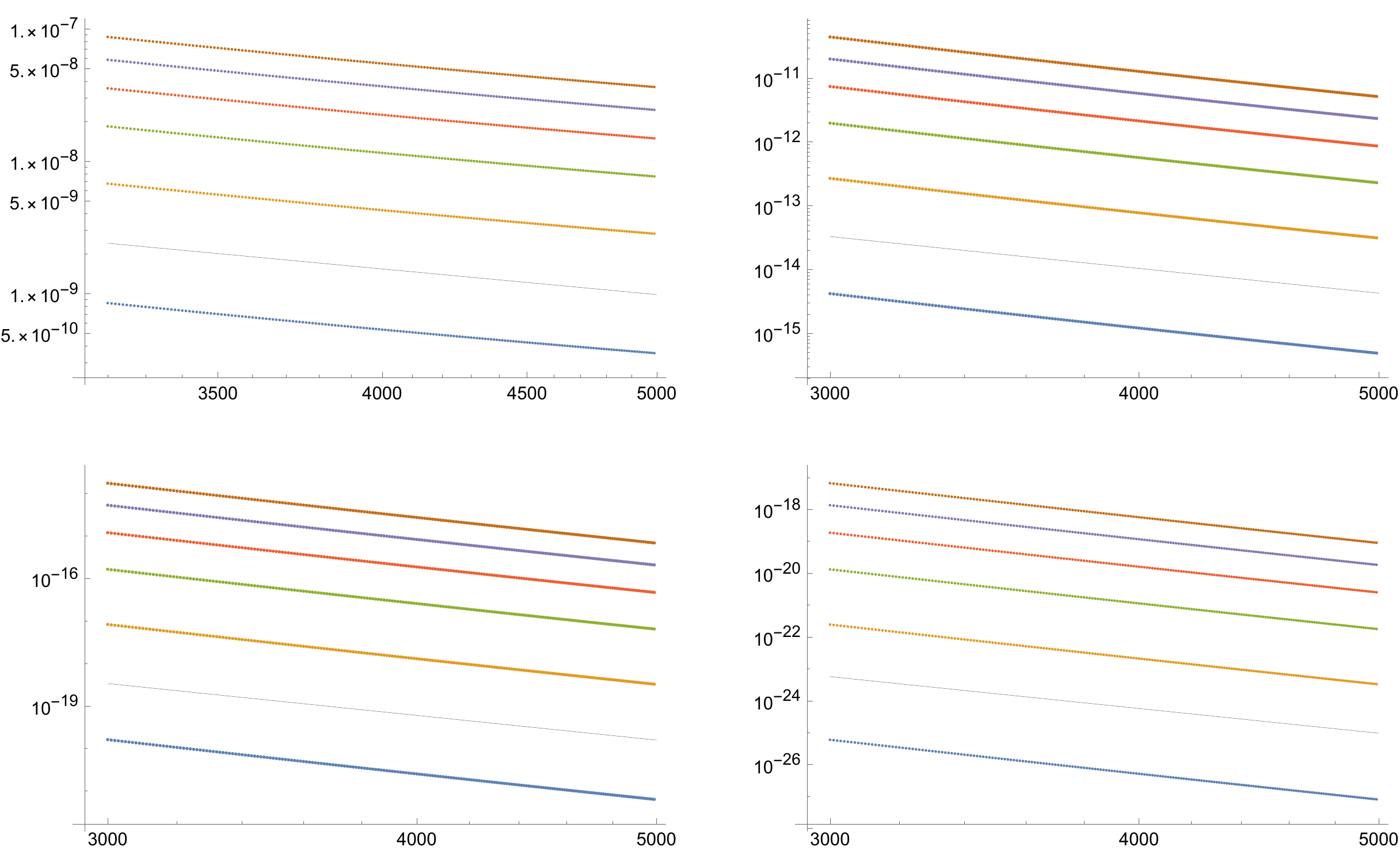}}
\put(23,54.5){$\ell=0$}
\put(70,54.5){$\ell=1$}
\put(23,27){$\ell=2$}
\put(70,27){$\ell=3$}
\put(88,1.75){$N$}
\put(88,30){$N$}
\put(42.75,1.75){$N$}
\put(42.75,30){$N$}
\put(2,54.5){$\frac{\Delta S_{0,\textrm{dS}}}{\Delta N}$}
\put(47.5,54.5){$\frac{\Delta S_{1,\textrm{dS}}}{\Delta N}$}
\put(2,26.5){$\frac{\Delta S_{2,\textrm{dS}}}{\Delta N}$}
\put(47.5,26.5){$\frac{\Delta S_{3,\textrm{dS}}}{\Delta N}$}
\put(91.5,21){\includegraphics[angle=0,width=0.085\textwidth]{dS_legendV}}
\end{picture}
\vspace{-0.7cm}
\caption{The finite difference $({S_{\ell,\textrm{dS}}(n,\tau,N+10)-S_{\ell,\textrm{dS}}(n,\tau,N)})/{10}$ for various values of $n$, $\tau=-3000$, and $\ell=0$ to $\ell=3$. The slopes are approximately $-1.952$, $-4.214$,$-6.356$ and $-8.451$ for $\ell=0$ to $\ell=3$ respectively. The solid gray lines have slope equal to $-2(\ell+1)$.  All lines of different $n$ exhibit similar behaviour.}
\label{fig:dS_3d_finite_t3000}
\centering
\begin{picture}(100,58.5)
\put(0,0){\includegraphics[angle=0,width=0.88\textwidth]{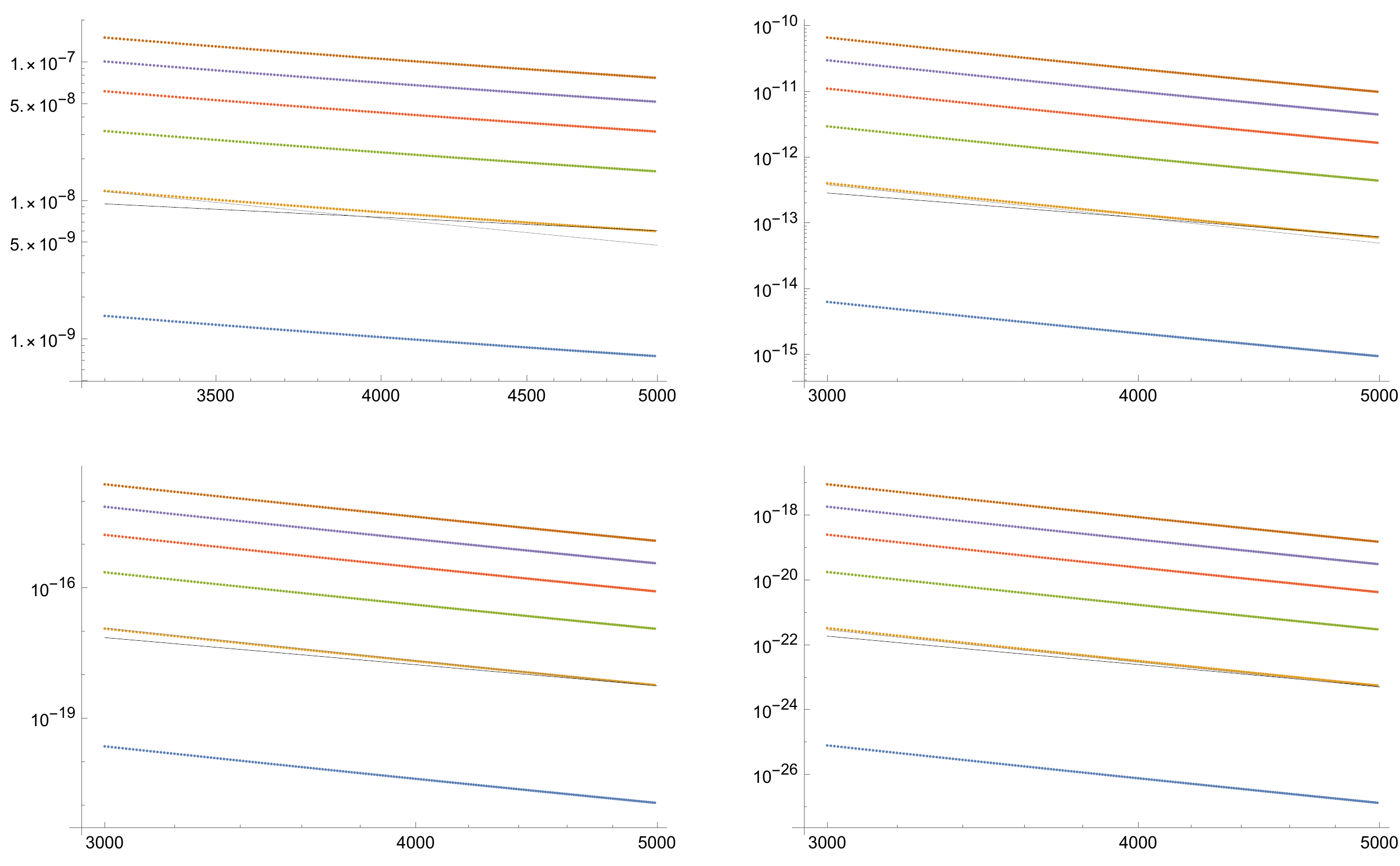}}
\put(23,54.5){$\ell=0$}
\put(70,54.5){$\ell=1$}
\put(23,27){$\ell=2$}
\put(70,27){$\ell=3$}
\put(88,1.75){$N$}
\put(88,30){$N$}
\put(42.75,1.75){$N$}
\put(42.75,30){$N$}
\put(2,54.5){$\frac{\Delta S_{0,\textrm{dS}}}{\Delta N}$}
\put(47.5,54.5){$\frac{\Delta S_{1,\textrm{dS}}}{\Delta N}$}
\put(2,26.5){$\frac{\Delta S_{2,\textrm{dS}}}{\Delta N}$}
\put(47.5,26.5){$\frac{\Delta S_{3,\textrm{dS}}}{\Delta N}$}
\put(91.5,21){\includegraphics[angle=0,width=0.085\textwidth]{dS_legendV}}
\end{picture}
\vspace{-0.7cm}
\caption{The finite difference $({S_{\ell,\textrm{dS}}(n,\tau,N+10)-S_{\ell,\textrm{dS}}(n,\tau,N)})/{10}$ for various values of $n$, $\tau=-1800$, and $\ell=0$ to $\ell=3$. The slopes are approximately $-1.497$, $-3.721$,$-5.870$ and $-7.999$ for $\ell=0$ to $\ell=3$, respectively. The solid gray lines red have slope equal to $-2(\ell+1)$, while the black ones $-(2\ell+1)$.  All lines of different $n$ exhibit similar behaviour.}
\label{fig:dS_3d_finite_t1800}
\end{figure*}

\begin{figure*}[t]
\centering
\begin{picture}(100,57.5)
\put(0,0){\includegraphics[angle=0,width=0.88\textwidth]{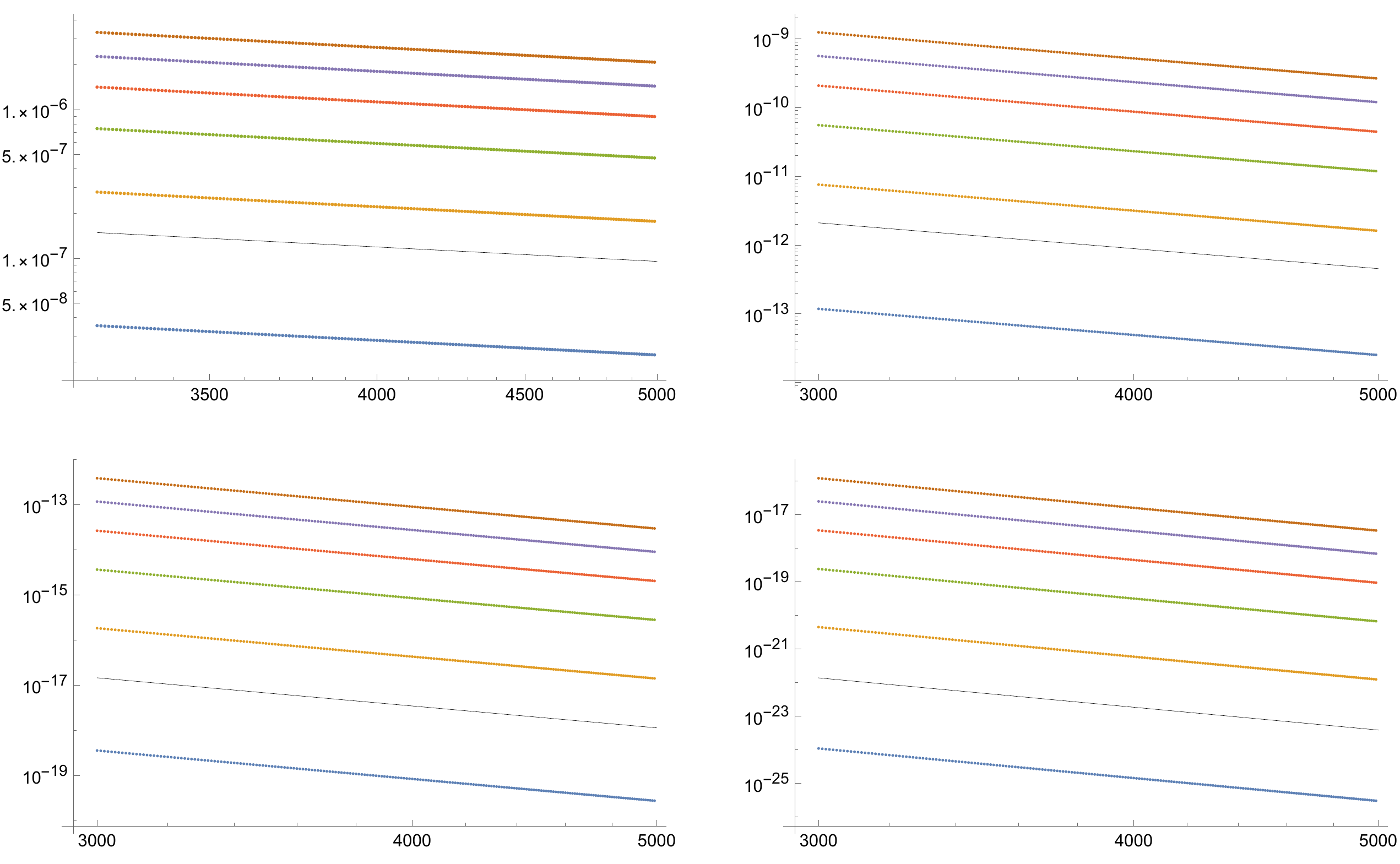}}
\put(23,54.5){$\ell=0$}
\put(70,54.5){$\ell=1$}
\put(23,27){$\ell=2$}
\put(70,27){$\ell=3$}
\put(88,1.75){$N$}
\put(88,30){$N$}
\put(42.75,1.75){$N$}
\put(42.75,30){$N$}
\put(2,54.5){$\frac{\Delta S_{0,\textrm{dS}}}{\Delta N}$}
\put(47.5,54.5){$\frac{\Delta S_{1,\textrm{dS}}}{\Delta N}$}
\put(2,26.5){$\frac{\Delta S_{2,\textrm{dS}}}{\Delta N}$}
\put(47.5,26.5){$\frac{\Delta S_{3,\textrm{dS}}}{\Delta N}$}
\put(91.5,21){\includegraphics[angle=0,width=0.085\textwidth]{dS_legendV}}
\end{picture}
\caption{The finite difference $({S_{\ell,\textrm{dS}}(n,\tau,N+10)-S_{\ell,\textrm{dS}}(n,\tau,N)})/{10}$ for various values of $n$, $\tau=-300$ and $\ell=0$ to $\ell=3$. The slopes are approximately $-1.037$, $-3.027$,$-5.036$ and $-7.044$ for $\ell=0$ to $\ell=3$, respectively. The solid black lines  have slope equal to $-(2\ell+1)$. All lines of different $n$ exhibit similar behaviour.}
\label{fig:dS_3d_finite_t300}
\end{figure*}

Using finite differences, we examine three indicative instants in time for the $\ell=0$ to $\ell=3$ sectors. In Figures \ref{fig:dS_3d_finite_t3000}, \ref{fig:dS_3d_finite_t1800} and \ref{fig:dS_3d_finite_t300} we present the rate of change of $S_{\ell,\textrm{dS}}(n,\tau,N)$ with respect to $N$ and estimate its slope for $\tau=-3000,\tau=-1800$ and $\tau=-300$ respectively. We already know from the vacuum analysis that, for $\tau\rightarrow-\infty$, the slope is approximately $-(2\ell+3)$. We notice in Figure \ref{fig:dS_3d_finite_t3000}, for $\tau\simeq-3000$, that the slope shifts to approximately $-2(\ell+1)$ and keeps shifting. Finally, for $\tau\simeq-300$, it becomes $-(2\ell+1)$, as is evident from Figure \ref{fig:dS_3d_finite_t300}. It must be kept in mind that this slope identifies the dominant finite-size effects. The entanglement entropy always includes a finite-size term of the form $N^{-2\ell}$, but whether this is dominant or not depends on time.

The above imply that gradually the background evolution induces the transition from equation \eqref{eq:vac_3d_N_corrections} to \eqref{eq:dS_3d_N_corrections}. In particular, the  $\ell=0$ sector of the theory develops a logarithmic dependence on the size of the \textit{overall system}, as parameterized by $N$. In the continuum limit, this sector and the $(1+1)$-dimensional toy model coincide. As a result, this $N$-dependence is the same as that discussed in Appendix \ref{sec:1d_dS}. The contributions to the entropy from the  remaining angular-momentum sectors tend to a constant value for $N\rightarrow\infty$. However, the scaling with $N$ of the corrections is shifted by a power of $2$.

Apart from analyzing the $\ln N$ term, our goal is to isolate the $N$-independent part of the entanglement entropy and study its behaviour as a function of $n$ and $\tau$. Thus, we repeat the process described in appendices \ref{sec:3d_flat} and \ref{sec:1d_dS}, albeit with slight modifications. First we repeat the calculation of the entanglement entropy $S_{\ell,\textrm{dS}}(n,\tau,N_0)$ for every angular-momentum sector, for a convenient value $N_0=60$ and $1\leq n\leq 59$, up to $\ell_\textrm{max}=10^6$. The calculation must be performed for several instants in time. More specifically, we analyze $55$ instants between $\tau=-3000$ and $\tau=-300$ in steps of $50$.

The next step requires the subtraction of the finite-size effects. As in the flat-space case in Appendix \ref{sec:3d_flat}, this is done for $\ell\leq 220$, since, for larger angular momenta, the corrections are insignificant compared to the numerical precision. For every $\ell$-sector, we extract the asymptotic  value of $S_{\ell,\textrm{dS},\infty}(n,\tau)$ for every $n$ and $\tau$ by fitting through equation \eqref{eq:dS_3d_N_corrections}, allowing for $10$ subleading corrections in $1/N$ for the most important $\ell=0$ to $\ell=4$ sectors, and $3$ subleading corrections for the rest. The most important deviations from the flat-space case occur in the $\ell=0$ sector, where we must include the logarithmic term $\ln N$ in the fit, and isolate its coefficient for every $n$ and $\tau$.

Subsequently, we extract the $\ell_\textrm{max}\rightarrow\infty$ limit, as in the flat-space case. After a thorough review of the truncated sum for every instant in time, we conclude that its scaling behaviour with respect to $\ell_\textrm{max}$ is the same as in flat space, see equation \eqref{eq:lmax_exp}. Thus, we assume that the truncated sums are described by \eqref{eq:lmax_exp_dS}. To prove that the leading term scales as $\ell_\textrm{max}^{-2}$, we calculate the rate of change of $S_{\textrm{dS},\infty}\left(n,\tau;\ell_{\textrm{max}}\right)$ with respect to $\ell_{\textrm{max}}$ for fixed $n$ and $\tau$. It is apparent from Figure \ref{fig:lmax_limit_dS_t300} that this rate of change is proportional to $\ell_\textrm{max}^{-3}$. However, equation \eqref{eq:lmax_exp_dS} indicates that subleading terms, both polynomial and logarithmic, must also be included. By fitting with $10$ polynomial and $10$ logarithmic corrections, we extract the asymptotic value of the sum for $\ell_\textrm{max}\rightarrow\infty$ for every $n$ and $\tau$.

\begin{figure*}[t]
\centering
\begin{picture}(60,37.25)(0,0)
\put(0,0){\includegraphics[width=0.55\textwidth]{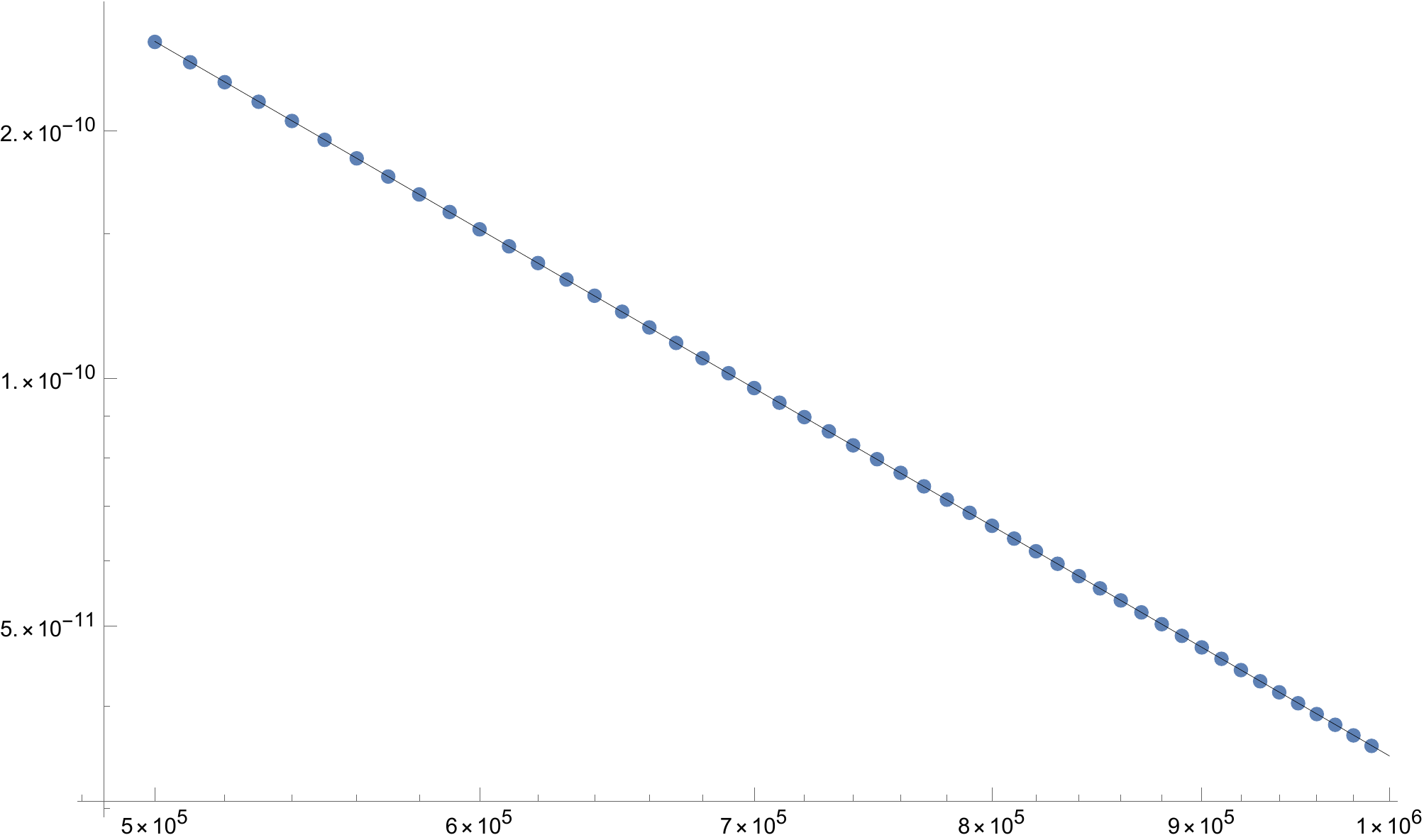}}
\put(55.5,1){$\ell_{\textrm{max}}$}
\put(0,33.75){$\frac{\Delta S_{\textrm{dS},\infty}}{\Delta\ell_{\textrm{max}}}$}
\end{picture}
\caption{The finite difference $({S_{\textrm{dS},\infty}(n,\tau,\ell_\textrm{max}+10^4)-S_{\textrm{dS},\infty}(n,\tau,\ell_\textrm{max})})/{10^4}$ for $n=50,\tau=-300$ and various values of $\ell_\textrm{max}$. A linear fit verifies that the slope is approximately $-2.885$. The small deviation from the value $3$ is due to subleading corrections.}
\label{fig:lmax_limit_dS_t300}
\end{figure*}
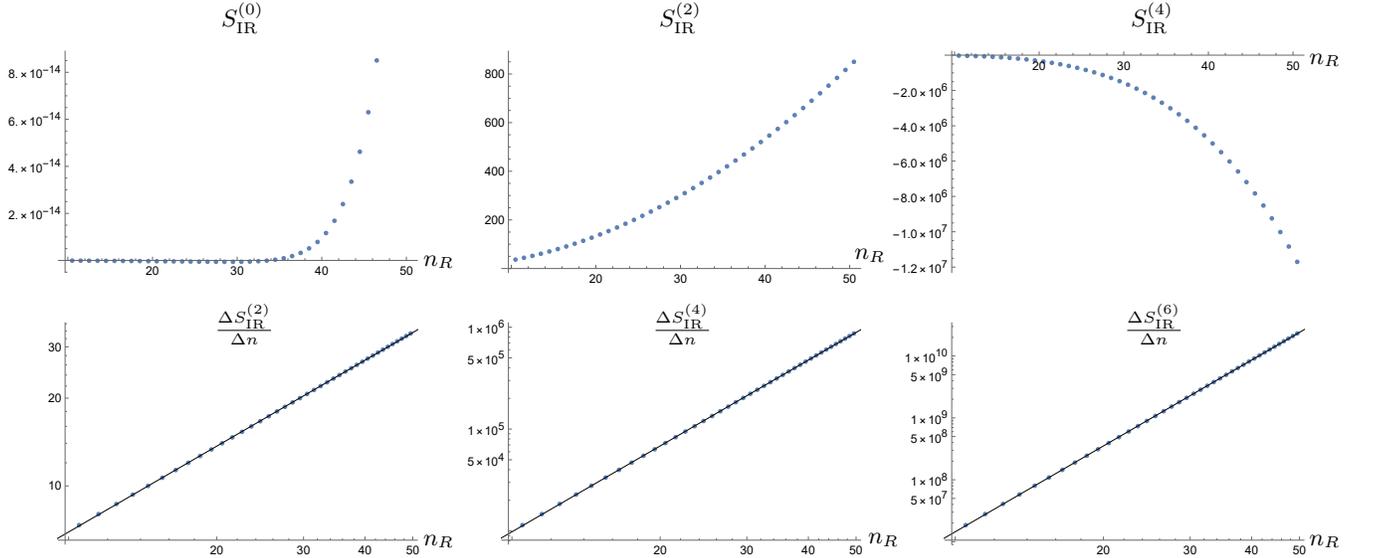
\begin{figure*}[t]
\centering
\begin{picture}(100,43)
\put(0,0){\includegraphics[angle=0,width=0.965\textwidth]{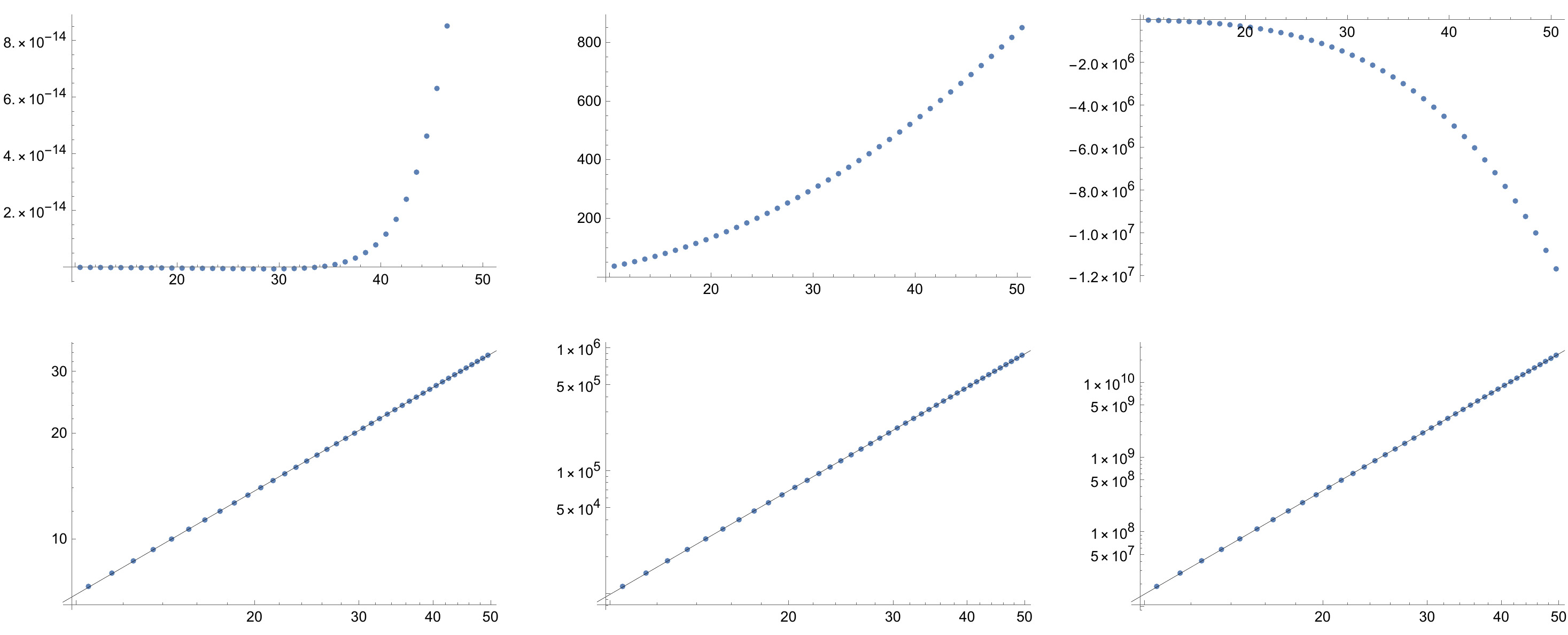}}
\put(16,40){$S_{\textrm{IR}}^{(0)}$}
\put(48.5,40){$S_{\textrm{IR}}^{(2)}$}
\put(83.5,40){$S_{\textrm{IR}}^{(4)}$}
\put(15.5,17){$\frac{\Delta S_{\textrm{IR}}^{(2)}}{\Delta n}$}
\put(48,17){$\frac{\Delta S_{\textrm{IR}}^{(4)}}{\Delta n}$}
\put(83,17){$\frac{\Delta S_{\textrm{IR}}^{(6)}}{\Delta n}$}
\put(31,1.5){$n_R$}
\put(64,1.6){$n_R$}
\put(96.75,1.5){$n_R$}
\put(31,22.25){$n_R$}
\put(63,22.75){$n_R$}
\put(96.75,37.25){$n_R$}
\end{picture}
\caption{The first row depicts $S_{\textrm{IR}}^{(0)}(n)$ (left panel), $S_{\textrm{IR}}^{(2)}(n)$ (middle panel) and $S_{\textrm{IR}}^{(4)}(n)$ (right panel). $S_{\textrm{IR}}^{(0)}(n)$ is of very small amplitude and must be considered identically zero, in accordance with the fact that no such term is expected in the flat space limit. The second row depicts the finite differences $\vert S_{\textrm{IR}}^{(2)}(n+1) -S_{\textrm{IR}}^{(2)}(n) \vert$ (left panel), $\vert S_{\textrm{IR}}^{(4)}(n+1) -S_{\textrm{IR}}^{(4)}(n) \vert$ (middle panel) and $\vert S_{\textrm{IR}}^{(6)}(n+1) - S_{\textrm{IR}}^{(6)}(n) \vert$ (right panel). The solid black lines are linear fits having slope $0.98$, $2.80$ and $4.61$, implying that the leading behaviour of the corrections are of the form $\left(n_R/\tau\right)^{2i}$. Recall that $n_R=n+1/2$ and $R=n\epsilon$ in the continuous theory.}
\label{fig:dS_3d_IR_terms}
\end{figure*}

After the above manipulations, the data are finally in a form that can be utilized in order to study the continuous theory. In contrast to the flat case, in de Sitter space the entanglement entropy does not depend only on the dimensionless parameter $n$, but also on the conformal time $\tau$ and the dimensionless size $N$ of the \textit{overall system}. This last parameter affects only the $\ell=0$ sector. The entropy takes the form
\begin{equation}
S_\textrm{dS}(n,\tau,N)=S_\textrm{IR}(n,\tau)\ln N + S_{\textrm{dS},\infty}(n,\tau).
\end{equation}
In the following we study the behaviour of $S_\textrm{IR}(n,\tau)$ and $S_{\textrm{dS},\infty}(n,\tau)$.

\begin{figure*}[ht]
	\centering
	\begin{picture}(98,34)
		\put(2,0){\includegraphics[angle=0,width=0.93\textwidth]{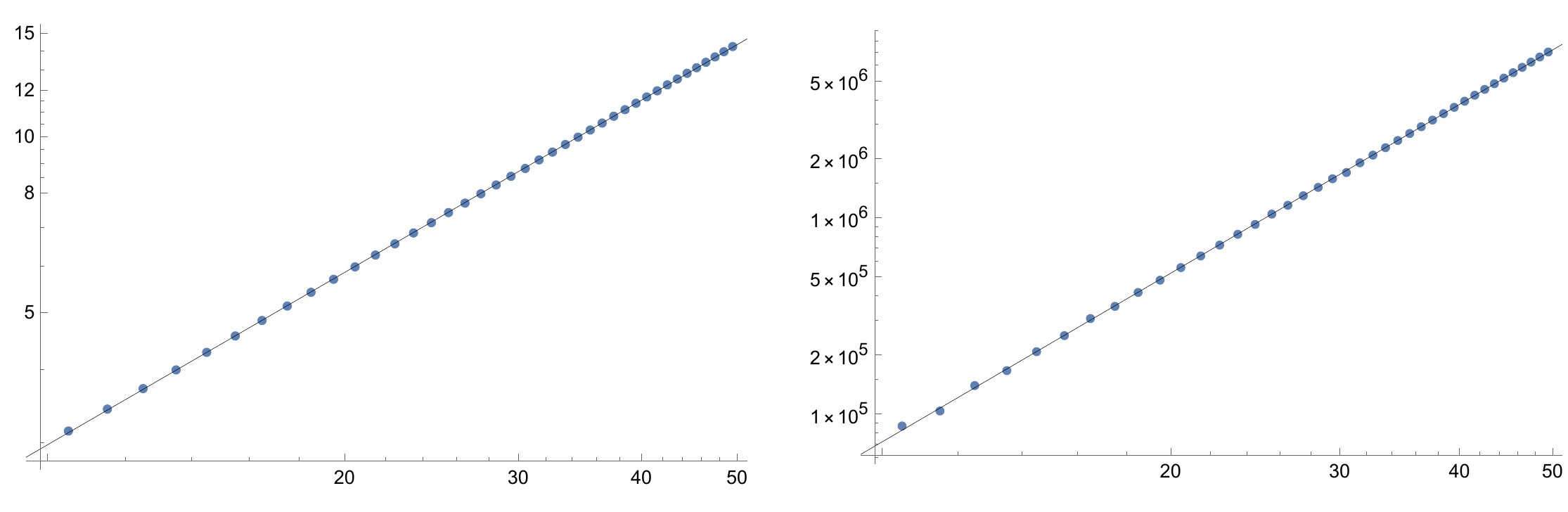}}
		\put(46.5,2.75){$n_R$}
		\put(49.5,30){$\frac{\Delta S_{\textrm{dS},\infty}^{(4)}}{\Delta n}$}
		\put(95,3){$n_R$}
		\put(0,30){$\frac{\Delta S_{\textrm{dS},\infty}^{(2)}}{\Delta n}$}
	\end{picture}
	\caption{The finite difference $\vert S_{\textrm{dS},\infty}^{(2)}(n+1) -S_{\textrm{dS},\infty}^{(2)}(n) \vert$ (left panel) and  $\vert S_{\textrm{dS},\infty}^{(4)}(n+1) -S_{\textrm{dS},\infty}^{(4)}(n) \vert$ (left panel). The solid black lines have slope $0.98$ and $2.86$, respectively. Recall that $n_R=n+1/2$ and $R=n\epsilon$ in the continuous theory.}
	\label{fig:dS_3d_finite_terms}
\end{figure*}

For the relatively early times that we consider, the effects arising from the de Sitter background can be {treated as small corrections to the flat-space result. Thus, we can fit $S_\textrm{IR}(n,\tau)$ with a series like the one in equation \eqref{eq:tau_series_IR}. After the time fit is performed for every $n$ with terms up to $\tau^{-18}$, we plot the coefficients of $\tau^0,\tau^{-2}$ and $\tau^{-4}$ with respect to $n_R=n+1/2$ in the first row of Figure \ref{fig:dS_3d_IR_terms}.  As is evident and also expected from our previous vacuum analysis, there is no $\tau^0$ term, and the leading IR divergence scales with $\tau^{-2}$. In order to extract the scaling behaviour of subleading terms of the form $S_{\textrm{IR}}^{(2i)}(n)$ with respect to $n$, in the second row of Figure \ref{fig:dS_3d_IR_terms} we plot the finite difference $\vert S_{\textrm{IR}}^{(2i)}(n+1) -S_{\textrm{IR}}^{(2i)}(n) \vert$ for $i=1,2,3$. The solid black lines are linear fits with slopes $0.98,2.80$ and $4.61$, respectively, close to $1,3$ and $5$, implying that the leading behaviour of the corrections is $\left(n_R/\tau\right)^{2i}$, which becomes $\left(R/\tau\right)^{2i}$ in the continuous theory\footnote{Recall that for the discretized theory $\tau$ is measured in units of $\epsilon$.}. These are the only terms that survive in the continuum limit. Unfortunately, our numerical accuracy, albeit high, can only allow us to estimate the leading coefficient. By fitting with the expression
\begin{equation}
S_\textrm{IR}^{(2)}(n_R)=c_\textrm{IR}^{(2)}n_R^2+c_\textrm{IR}\ln n_R+d_\textrm{IR},
\end{equation}
we get
\begin{equation}
c_\textrm{IR}^{(2)}=0.3333364.
\end{equation}
In complete analogy with the $(1+1)$-dimensional theory, we are led to the conclusion that, for the specific time regime under consideration, the part of the entanglement entropy that depends on the size of the overall system behaves as
\begin{equation}
S_\textrm{IR}(R,\tau)=\frac{1}{3}\frac{H^2}{\tau^2}\ln\frac{L}{\epsilon}+\dots
\end{equation}

We next consider the $N$-independent part of the entanglement entropy $S_{\textrm{dS},\infty}(n,\tau)$. Once again, we are in  a regime where the data can be fit with a series like  \eqref{eq:tau_series_inf}. The term $S_{\textrm{dS},\infty}^{(0)}(n)$ represents the flat-space component of the entanglement entropy. By fitting again with the expression \eqref{eq:vac_exp}, we calculate the various coefficients. In table \ref{Tab:dS_3d_fit} we present our results. It is shown that the expected value of $d_1=-1/90$ can be once again derived with  high accuracy, this time from the de Sitter theory.

\begin{table}
\center
\begin{tabular}{|c | c | c| c |}
\hline
$i_\textrm{max}$ & $d_2$ & $d_0$ & $-1/d_1$\\
\hline
0 & 0.29543143401 & -0.035534 & 90.6016\\
\hline
1 & 0.29543144804 & -0.035396 & 90.2416\\
\hline
2 & 0.29543145267 & -0.035326 & 90.0779 \\
\hline
3 & 0.29543145351 & -0.035309 & 90.0403 \\
\hline
4 & 0.29543145385 & -0.035301 & 90.0218\\
\hline
5 & 0.29543145404 & -0.035295 & 90.0096\\
\hline
6 & 0.29543145413 & -0.035291 & 90.0009\\
\hline
7 & 0.29543145424 & -0.035287 & 89.9940\\
\hline
8 & 0.29543145428 & -0.035285 & 89.9895\\
\hline
\end{tabular}
\caption{The area-law coefficient $d_2$, the constant term $d_0$ and the denominator of the subleading logarithmic term of $S_{\textrm{dS},\infty}^{(0)}=d_2n_R^2+d_1\ln n_R+d_0$ for a varying number of subleading corrections. We verify the known value $d_1=-1/90$ with high accuracy. \label{Tab:dS_3d_fit}}
\end{table}

We now focus on $S_{\textrm{dS},\infty}^{(2)}(n)$ and $S_{\textrm{dS},\infty}^{(4)}(n)$. In Figure \ref{fig:dS_3d_finite_terms} we depict their rate of change with respect to $n$, determined to be $0.98$ and $2.86$ respectively. Our numerical precision allows us only to study the first term. Subsequent fits and the fact that the slope of $S_{\textrm{dS},\infty}^{(2)}$ is smaller than $1$ indicate that either there is no term $n_R^2\ln n_R$, in direct contrast with the $(1+1)$-dimensional analogue, or the accuracy of our data is not high enough to isolate it. More specifically, by performing a fit of the form
\begin{multline}\label{eq:finite_3d_fit}
S_{\textrm{dS},\infty}^{(2)}(n) = a^{(2)\prime}_{2} n_R^2\ln n_R + a^{(2)}_2 n_R^2 + c \ln n_R  \\ 
+ d+ \frac{a_{-2}}{n_R^2}+ \frac{a_{-4}}{n_R^4}+ \frac{a_{-6}}{n_R^6},
\end{multline}
we obtain
\begin{equation}
 a^{(2)\prime}_{2}=0.000025,\quad a^{(2)}_2=-0.142650.
\end{equation}

The difference between the $(1+1)$- and the $(3+1)$-dimensional theory regarding the fact that no $n_R^2\ln n_R$ appears in the latter, was intriguing enough to merit further investigation. By isolating only the $\ell=0$ contribution to $S_{\textrm{dS},\infty}^{(2)}(n)$, which is essentially identical to the $(1+1)$-dimensional problem, we find that this contribution results in a significant coefficient that equals $-0.333498$,  very close to} $-1/3$. However, all other angular-momentum sectors also contribute to the same term in $S_{\textrm{dS},\infty}^{(2)}(n)$ with a coefficient equal to $0.347750$, almost the opposite to that of the $\ell=0$ sector. Thus, it seems that the $\ell$-sectors for $\ell\geq 1$ cancel this certain scaling property of the $\ell=0$ sector and  the term $n^2_r\ln n_r$ is only a feature of the $(1+1)$-dimensional theory.

\bibliography{spooky}

\end{document}